\documentclass[a4paper,11pt]{article}
\pdfoutput=1

\usepackage{jheppub} 
                    
\usepackage[T1]{fontenc} 
\usepackage[dvipsnames]{xcolor}
\usepackage{hyperref}
\usepackage[capitalise]{cleveref}
\usepackage{lscape}
\usepackage{caption}
\usepackage{subcaption}
\usepackage{empheq}

\author[]{Jens Erler and}
\author[]{Rodolfo Ferro-Hernández}

\affiliation[]{
PRISMA+ Cluster of Excellence and  Institute for Nuclear Physics
Johannes Gutenberg University, 55099 Mainz, Germany.}

\emailAdd{: erler@uni-mainz.de}
\emailAdd{rferrohe@uni-mainz.de}

\abstract{

We compute a theoretically driven prediction for the hadronic contribution to the electromagnetic running coupling at the $Z$ scale using lattice QCD and state-of-the-art perturbative QCD. We obtain$$\Delta\alpha^{(5)}(M^2_Z)=\left[279.5\pm0.9\pm0.59\right]\times10^{-4}\quad\quad\,\,\,\,\,\,(\mathrm{Mainz \,\,\,Collaboration})$$$$\Delta\alpha^{(5)}(M^2_Z)=\left[278.42\pm0.22\pm0.59\right]\times10^{-4}\,\,\,\,\,\,\,\,\quad(\mathrm{ BMW \,\,\,Collaboration}),$$
 where the first error is the quoted lattice uncertainty. The second is due to perturbative QCD, and is dominated by the   parametric uncertainty on $\hat{\alpha}_s$, which is based on a rather conservative error. Using instead the PDG average, we find a total error on $\Delta\alpha^{(5)}(M^2_Z)$ of $0.4\times10^{-4}$. Furthermore, with a particular emphasis on the charm quark contributions, we also update $\Delta\alpha^{(5)}(M^2_Z)$  when low-energy cross-section data is used as an input, obtaining $\Delta\alpha^{(5)}(M^2_Z) = \left[276.29 \pm 0.38 \pm 0.62\right] \times 10^{-4}$. The difference between lattice QCD and cross-section-driven results reflects the known tension between both methods in the computation of the anomalous magnetic moment of the muon. Our results are expressed in a way that will allow straightforward modifications  and an easy implementation in electroweak global fits.

}

\begin{document}

\title{Perturbative contributions to  $\Delta\alpha^{(5)}(M^2_Z)$ }

\date{\today}

\maketitle

\section{\label{sec:level1} Introduction}
The fine-structure constant $\alpha$, a central parameter of the Standard Model, is essential for computing a wide range of observables and consistency relations. At low energy, it can be determined from atom interferometry in Cs \cite{Parker_2018} and Rb \cite{Morel:2020dww} atoms or from the anomalous magnetic moment of the electron \cite{PhysRevLett.100.120801,atoms7010028} with a relative error of approximately $\delta\alpha/\alpha \sim 10^{-10}$. At high energies, radiative corrections can be absorbed into the running electromagnetic coupling. The current level of relative error at $M_Z$ is $\delta\alpha/\alpha(M^2_Z) \sim 10^{-4}$, which will be insufficient for future colliders \cite{Proceedings:2019vxr}. The significant difference in uncertainties at the two scales arises from the hadronic and perturbative QCD contributions to the vacuum polarization function of the photon $\Pi(q^2)$.

The hadronic contributions can be estimated using $e^+e^- \rightarrow \mathrm{hadrons}$ data \cite{Davier:1998si, Davier:2010nc, Bodenstein:2012pw, Keshavarzi:2018mgv, Davier:2019can, Keshavarzi:2019abf, Jegerlehner:2019lxt} and the optical theorem, or by a direct calculation in a discretized Euclidean space (lattice QCD) \cite{Budapest-Marseille-Wuppertal:2017okr, Borsanyi:2020mff, Ce:2022eix}. Perturbative QCD (pQCD) can be applied at momentum transfer  $\sim 4\,\mathrm{GeV}^2$ and above. The contribution of all these QCD effects to $\alpha(M^2_Z)$ are denoted by $\Delta\alpha^{(5)}(M^2_Z)$.

The same information used to compute $\Delta\alpha^{(5)}(M^2_Z)$ is also used to calculate the anomalous magnetic moment of the muon $a_\mu$. The prediction of $a_\mu$ from cross-section data \cite{Davier:2019can, Keshavarzi:2019abf, Jegerlehner:2019lxt, Benayoun:2019zwh} is smaller and strongly conflicts with its experimental value \cite{Muong-2:2006rrc, Muong-2:2021ojo}, while lattice QCD \cite{Borsanyi:2020mff} results also predict a lower value but with less tension. With some caveats, a larger value of $a_\mu$ leads to a larger value of $\Delta\alpha^{(5)}(M^2_Z)$. On the other hand, in electroweak precision fits, $\Delta\alpha^{(5)}(M^2_Z)$ itself is used to compute the prediction of the $W$ mass, where an increase in the value of $\Delta\alpha^{(5)}(M^2_Z)$ results in a decrease in the prediction for $M_W$, while the experimental value is larger than the current SM prediction \cite{Erler:2022ynf}. Therefore, a reduction in the tension related to $a_\mu$ would likely lead to an increased tension in $M_W$. An account of this issue was given in Ref. \cite{Crivellin:2020zul} with a counter-reply in Ref. \cite{Borsanyi:2020mff}. Our results simplify how the implications of low energy lattice results can be studied in electroweak global fits.

We compute $\Delta\alpha^{(5)}(M^2_Z)$ using the renormalization group equations (RGE) \cite{Baikov:2012zm}, matching conditions \cite{Sturm:2014nva}, and the lattice QCD results from Ref.~\cite{Ce:2022eix} and Refs.~\cite{Budapest-Marseille-Wuppertal:2017okr, Borsanyi:2020mff} as input. Additionally, we provide an updated value  using low-energy cross-section data as input and a brief overview and comparison of the methods used by the data-driven collaborations in Refs.~\cite{Proceedings:2019vxr, Davier:2019can, Keshavarzi:2019abf, Erler:1998sy}. Alongside this, in \cref{LatticeRGE}  we provide a simple equation to relate the lattice QCD results to the low-energy integrals of $e^+e^- \rightarrow \mathrm{hadrons}$ data typically quoted in the literature.

There are computational differences between collaborations that calculate hadronic contributions to the electromagnetic running coupling from cross-section data. Each collaboration analyzes and models data at low energies differently or implements perturbative contributions in varying ways. Despite such differences, the reported results are in agreement. However, there is the possibility that a substantial portion of the error is parametric in nature, potentially obscuring tensions between different analyses. We find that the main difference between the explicit integration of $e^+e^- \rightarrow \mathrm{hadrons}$ data and the use of RGE plus matching conditions lies in the treatment of the charm quark, resulting in a smaller error for the latter.

Our results are presented in a form, which will facilitate the identification of areas for improvement to achieve a more precise extraction of $\alpha$. These expressions can serve as inputs for global fit implementations, enabling the incorporation of correlations between $a_{\mu}$, $\Delta\alpha^{(5)}(M^2_Z)$, and the running weak mixing angle \cite{Erler:2017knj} in a straightforward manner. For a study on how correlations between these parameters impact electroweak global fits see Refs.~\cite{Malaescu:2020zuc, Erler:2000nx, Erler:2022ynf}.

\section{Vacuum polarization function}

The vacuum polarization function can be defined using the integral expression,
\begin{equation}
\left(-q^2 \eta^{\mu\nu}+q^{\mu} q^{\nu} \right)\hat{\Pi}(q^2,\mu^2)=i\int d^4x e^{iqx}\langle 0|TJ_{em}^{\mu}(x)J_{em}^{\nu}(0)| 0\rangle,
\end{equation}
where  $\mu$ appears due to the renormalization condition imposed by the $\overline{\mathrm{MS}}$ scheme and  $J^{\nu}_{em}$ is the electromagnetic current.  On the other hand, the on-shell  or substracted vacuum polarization, denoted by $\Pi(q^2)$ \textit{i.e} without the caret symbol, satisfies $\Pi(0)=0$. It follows that, \begin{equation}
    \Pi(q^2)=\hat{\Pi}(q^2,\mu^2)-\hat{\Pi}(0,\mu^2).\label{eq:vacuumbothschemes}
\end{equation}
The effective coupling is constructed to absorb the large logarithms that appear in this expression and is given by \begin{equation}
    \alpha(q^2)=\frac{\alpha}{1-\Delta \alpha (q^2)},\quad\quad\Delta \alpha (q^2)\equiv -4\pi \alpha \mathrm{Re}\, \left[\Pi(q^2)\right],\label{onshelldef}
\end{equation} where $\alpha$ is the fine structure constant in the Thomson limit.
In the $\overline{\mathrm{MS}}$ scheme, the running of $\hat{\alpha}$ is given by \cite{Sirlin:1980nh,Degrassi:2003rw} \begin{equation}
    \hat{\alpha}(\mu^2)=\frac{\alpha}{1-\Delta \hat{\alpha} (\mu^2)} \quad\quad \Delta \hat{\alpha} (\mu^2)\equiv 4\pi \alpha\hat{\Pi}(0,\mu^2).\label{msdef}
\end{equation}
Up to small corrections, the vacuum polarization can be split into a leptonic ($\hat{\Pi}_\mathrm{lep}(q^2,\mu^2)$) and an hadronic (QCD) piece ($\hat{\Pi}_\mathrm{had}(q^2,\mu^2)$), the leptonic piece is known up to four loops and its error is well under control \cite{Sturm:2013uka}. On the other hand, 
the size of the  perturbative corrections and non-perturbative effects make the QCD piece the dominant source of uncertainty on $\Delta\alpha$. We will focus on it in the following.  From \cref{eq:vacuumbothschemes,onshelldef,msdef}, 
\begin{equation}
    \Delta\alpha_{\mathrm{had}}(q^2)=\Delta \hat{\alpha}_{\mathrm{had}}(\mu^2)-4\pi\alpha \mathrm{Re}\,\left[\hat{\Pi}_{\mathrm{had}}(q^2,\mu^2)\right]. \label{change scheme}
\end{equation}
Note that \cref{change scheme} is the hadronic part of the relation between the on-shell and $\overline{\mathrm{MS}}$ definitions of the running coupling. 

The vacuum polarization function of  a heavy quark  can be written as a power series in the strong coupling constant $\hat{\alpha}_s$ (in the following we may also use the symbol $\hat{a}_s\equiv\frac{\hat{\alpha}_s}{\pi}$). The coefficients of this expansion are known exactly up to order $\hat{\alpha}^2_s$ including the full quark mass dependence \cite{Chetyrkin:1996cf,Maier:2007yn, Maier:2011jd}. At $\hat{\alpha}_s^3$ few terms of a power series in $z\equiv \frac{s}{4m^2_q}$,  $\sqrt{1-z}$ and $\frac{1}{z}$, are known, the so-called low energy, threshold  and high energy expansions respectively (see Refs.~\cite{Chetyrkin:2000zk,Chetyrkin:2006xg, Hoang:2001mm, Maier:2008he,Hoang:1998xf, Hoang:1997sj,Maier:2009fz}). In spite of this partial knowledge, there are techniques to approximately  reconstruct the mass dependence using the analytic properties of the vacuum polarization, through a conformal mapping and Padé approximants as shown in Refs.~\cite{Hoang:2008qy,Greynat:2010kx,Kiyo:2009gb,Greynat:2011zp} and  tested in Ref.~\cite{Maier:2017ypu}.

To handle light quarks, their contributions have to be first computed at a scale $\sqrt{s} \gtrsim 1.8 \mathrm{\,GeV}$ using nonperturbative methods like lattice QCD or experimental cross-section data. Once determined, contributions at a higher scale can be computed  perturbatively.
To use cross-section data, one exploits that $\hat{\Pi}_{\mathrm{had}}(s,\mu^2)$ is analytic in the $s$ complex plane except for poles and branch-cuts in the positive real axis.  $\hat{\Pi}_{\mathrm{had}}(q^2,\mu^2)$ can then be obtained from Cauchy's theorem,  \begin{equation}
    \hat{\Pi}_{\mathrm{had}}(q^2,\mu^2)=\frac{1}{2\pi i}\oint_C \frac{\hat{\Pi}_{\mathrm{had}}(s,\mu^2)}{s-q^2}ds,
\end{equation} 
\begin{figure*}
    \centering
    \includegraphics[scale=0.25]{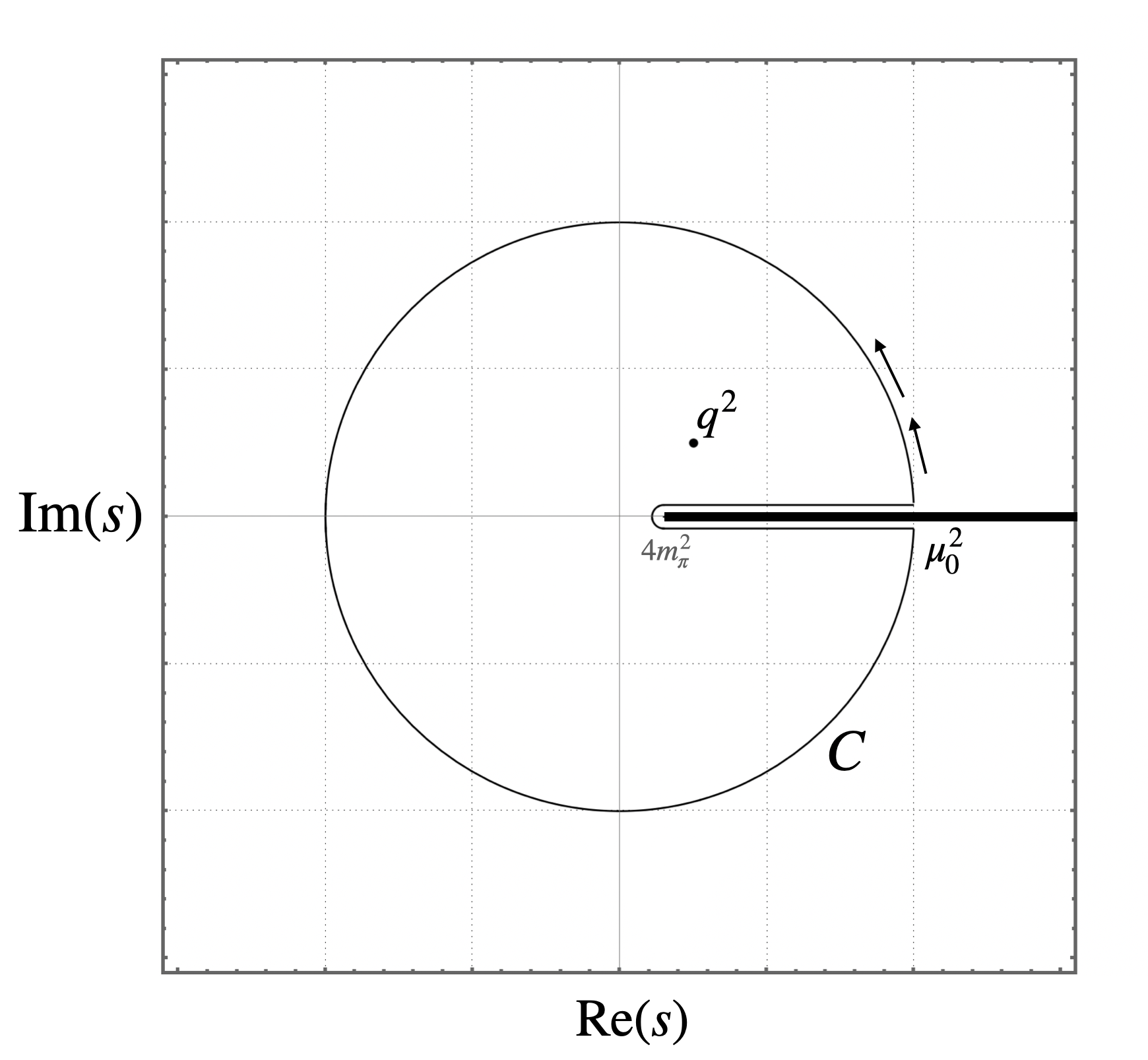}
    \caption{Region of integration to extract the vacuum polarization function at some point $q^2$ inside $C$.}
    \label{fig:ContourRegion}
\end{figure*}where $C$ is the contour shown in \cref{fig:ContourRegion}, and $q^2$ is a point inside such a contour.
The next step is to split this contour into an integral along the positive real axis, and an integral over a circle. The integral along the real axis picks up the discontinuity of the vacuum polarization function when evaluated above and below the positive real axis, which is proportional to $\mathrm{Im}\, \hat{\Pi}$. Then using the optical theorem one gets
\begin{equation}
    \hat{\Pi}_{\mathrm{had}}(q^2,\mu^2)=\frac{1}{12\pi^2}\int^{\mu^2_0}_{4m^2_\pi}\frac{R(s)}{s-q^2-i\epsilon}ds+\frac{1}{2\pi i}\int_{|s|=\mu^2_0} \frac{\hat{\Pi}_{\mathrm{had}}(s,\mu^2)}{s-q^2}ds,\label{eq:master}
\end{equation} 
where $R(s)=12\pi \mathrm{Im}\,\hat{\Pi}_{\mathrm{had}}(s,\mu)=12\pi \mathrm{Im}\,\Pi_{\mathrm{had}}(s)$ is the ratio of the cross section of $e^+e^-\rightarrow\mathrm{hadrons}$ over $e^+e^-\rightarrow\mu^+\mu^-$.  \cref{eq:master} and \cref{eq:vacuumbothschemes} imply,
\begin{equation}
{\Pi_{\mathrm{had}}(q^2)}=\frac{q^2}{12\pi^2}\int^{\mu^2_0}_{4m^2_\pi}\frac{R(s)}{s(s-q^2-i\epsilon)}ds+\frac{q^2}{2\pi i}\int_{|s|=\mu^2_0} \frac{\hat{\Pi}_{\mathrm{had}}(s,\mu^2)}{s(s-q^2)}ds,\label{eq:onshellcontour}
\end{equation}
where in the limit $\mu_0\rightarrow\infty$ the second term on the right-hand side vanishes and one is left with, \begin{equation}
    \Pi_{\mathrm{had}}(q^2)=\frac{q^2}{12\pi^2}\int^{\infty}_{4m^2_\pi}\frac{R(s)}{s(s-q^2-i\epsilon)}ds.\label{eq:pifromR}
\end{equation} In terms of $\Delta\alpha$ this reads
\begin{equation}
    \Delta\alpha_{\mathrm{had}}(q^2)=-4\pi\alpha\,\mathrm{Re}\,\Pi_{\mathrm{had}}(q^2)=-\mathrm{Re}\left[\frac{\alpha q^2}{3\pi}\int^{\infty}_{4m^2_\pi}\frac{R(s)}{s(s-q^2-i\epsilon)}ds\right].
    \label{eq:onshellintegral}
\end{equation} 
In the following  we will employ a superscript to indicate the number of quarks considered. For example, $\Delta\alpha^{(5)}(M^2_Z)$ represents the contribution of 5 quarks $(u,d,s,c,b)$ to  $\alpha$ at $M_Z$.

\section{Explicit integration of $R$ in the timelike region}

One way to compute $\Delta\alpha^{(5)}(M^2_Z)$ is to use  \cref{eq:onshellintegral} with $q^2=M^2_Z$, which requires the knowledge of $R(s)$ for all $s$ on the positive real axis. By replacing the kernel function inside the integral, the same $R(s)$ can be used to compute the hadronic contribution to the anomalous magnetic moment of the muon $a_\mu$. Since the kernel of $a_\mu$ drops faster for large $s$ than that of $\Delta\alpha^{(5)}(M^2_Z)$,  $a_\mu$ has an enhanced sensitivity to low energy physics relative to $\Delta\alpha^{(5)}(M^2_Z)$.  In fact, the $\pi^+\pi^-$ channel gives around $70\%$ of the total hadronic contribution to $a_\mu$, compared to $10\%$ for $\Delta\alpha^{(5)}(M^2_Z)$.  The integral  in \cref{eq:onshellintegral} is usually divided into regions according to the reliability of data or pQCD values of $R$. This dichotomy is more relevant for $\Delta\alpha^{(5)}(M^2_Z)$ than for $a_\mu$ since pQCD can only be used at high energies.
The most recent results of $\Delta\alpha^{(5)}(M^2_Z)$ using this method are  DHMZ19 \cite{Davier:2019can}, KNT19 \cite{Keshavarzi:2019abf} and J19 \cite{Jegerlehner:2019lxt}. The collaborations differ by the choice of regions in which pQCD is used for $R(s)$. Above $1.8\,\mathrm{GeV}$, the main difference is that DHMZ19 uses pQCD everywhere except for the region between~$3.7-5\,\mathrm{GeV}$ whereas  KNT19 only uses it above $\sim11\,\mathrm{GeV}$. There are also significant differences in specific channels for data below $1.8\,\mathrm{GeV}$. A more comprehensive comparison between the two collaborations can be found in Ref.~\cite{Keshavarzi:2018mgv}.  

As an example, here we will recompute the perturbative contributions\footnote{We use   $\overline{\mathrm{MS}}$ masses as input parameters.} to $R$ in the same regions as DHMZ19. We will not recompute the data driven intervals, which are instead taken directly from DHMZ19. The data driven contributions to the integral as given in DHMZ19 are\footnote{Using the result for the quark hadron duality violations from DHMZ19 in the interval $1.8-2.0\,\mathrm{GeV}$ we are able to compute the data driven result at $2.0\,\mathrm{GeV}$ which is a standard point for lattice calculations.}  
\begin{equation}
    \frac{\alpha M^2_Z}{3\pi}\int^{(2\,\mathrm{GeV})^2}_{4m^2_\pi}\frac{R(s)}{s(M^2_Z-s)}ds=\left(58.71\pm0.09_\mathrm{stat}\pm0.29_{\mathrm{sys}}\pm0.23_\mathrm{csys}\pm0.28_{\mathrm{dual}}\right) \times10^{-4},\label{lowintegral}
\end{equation} 
and 
 \begin{equation}
    \frac{\alpha M^2_Z}{3\pi}\int^{(5.0\,\mathrm{GeV})^2}_{(3.7\,\mathrm{GeV})^2}\frac{R(s)}{s(M^2_Z-s)}ds=\left(15.79\pm0.12_\mathrm{stat}\pm0.66_{\mathrm{sys}}\right) \times10^{-4},
    \label{charmlowint}
\end{equation} 
while for  the $J/\psi$ and $\psi(2S)$ resonances the collaboration quotes \begin{equation}
 \frac{\alpha M^2_Z}{3\pi}\int\frac{R_{J/\psi}(s)}{s(M^2_Z-s)}ds=(7.00\pm0.13)\times 10^{-4},\quad \frac{\alpha M^2_Z}{3\pi}\int\frac{R_{\psi}(s)}{s(M^2_Z-s)}ds=(2.48\pm 0.08)\times10^{-4},\label{resonancesint}
\end{equation} using a Breit Wigner shape. The subscript "dual" denotes the error coming quark hadron duality violations as taken in DHMZ19, that is, as the difference between data and theory in the range $1.8-2\mathrm{\,GeV}$.  To compute $R$ perturbatively \cite{Baikov:2015tea,Baikov:2008jh,Chetyrkin:2000zk,Maier:2011jd,Chetyrkin:1996cf} we used the code\footnote{ We confirmed the results using our own implementation of $R$.} rhad \cite{Harlander:2002ur}. We obtain
\begin{align}
\Delta\alpha^{(5)}(M_Z)=&\left[275.77+ 136\,\Delta\hat{\alpha}_s+0.7\,\Delta \hat{m}_c-1.3\,\Delta\hat{m}_b\pm 0.67_{\mathrm{c-thr}}\pm 0.19_{\mathrm{tr}}\pm 0.28_{\mathrm{dual}}\right.\nonumber\\
          \pm&\left.0.38_{\mathrm{dat}<2\,\mathrm{GeV}}\pm0.15_{J/\psi}\right]\times10^{-4}\quad\quad\quad\mathrm{(Explicit \,\,integration\,\,method)},
          \label{eq:ExplicitIntegration}
\end{align}
which is in agreement with the number quoted in \cite{Davier:2019can}. In \cref{eq:ExplicitIntegration} we defined  \begin{equation}
            \Delta\hat{\alpha}_s\equiv \hat{\alpha}_s(M_Z)-0.1185\quad \Delta\hat{m}_c\equiv \hat{m}_c\left(\hat{m}_c\right)-1.270\,\mathrm{GeV}\quad\Delta\hat{m}_b\equiv \hat{m}_b\left(\hat{m}_b\right)-4.180\,\mathrm{GeV}.
            \end{equation}
The subscript "c-thr" denotes the error from the charm threshold region ${[3.7-5.0]\, \mathrm{GeV}}$. The subscript "tr" means truncation error, defined to be the size of the last known term in the perturbative series\footnote{In \cite{Davier:2019can}  also the difference between contour improved and fixed order perturbation theory is added to the error.}. This definition will allow for a simpler comparison between different frameworks to compute $\Delta\alpha^{(5)}(M^2_Z)$.  The subscript "$\mathrm{dat}<2 \,\mathrm{GeV}$" means the combined error from the low energy data in \cref{lowintegral} without the quark hadron duality error. Finally the subscript "$J/\psi$" denotes the error coming from the $J/\psi$ and $\psi(2S)$ resonances. When quoting final numerical results we will use as errors $\delta\hat{m}_c=\pm0.008\,\mathrm{GeV}$,  $\delta\hat{m}_b=\pm0.008\,\mathrm{GeV}$ and $\delta\hat{\alpha}_s(M_Z)=\pm0.016$ \cite{Workman:2022ynf}.

Since the threshold region of the charm quark gives the largest uncertainty, it is a good idea to isolate its contribution. The charm quark contribution in the threshold region is calculated by subtracting the pQCD light quark contribution, given by 
\begin{equation}
    \frac{\alpha M^2_Z}{3\pi}\int^{(5.0\,\mathrm{GeV})^2}_{(3.7\,\mathrm{GeV})^2}\frac{R^{\mathrm{(uds)}}(s)_{\mathrm{pert}}}{s(M^2_Z-s)}ds=\left[10.03+10\,\Delta\hat{\alpha}_s\pm0.03_{\mathrm{tr}}\right]\times 10^{-4},
    \label{eq:Timelikelight}
\end{equation}
while the charm quark contribution above $5\,\mathrm{GeV}$ can be computed explicitly within pQCD. 

Substracting  \cref{eq:Timelikelight} from  \cref{charmlowint}, adding \cref{resonancesint} and finally  adding the charm contribution to the continuum above $5\,\mathrm{GeV}$ we get \begin{equation}
{\Delta\alpha^{(c)}(M^2_Z)}=\bigg[78.72+27\,\Delta\hat{\alpha}_s+0.7\,\Delta \hat{m}_c\pm0.02_{\mathrm{tr}}\pm0.67_{\mathrm{c-thr}}\pm0.15_{J/\psi}\nonumber\bigg]\times10^4.\label{eq:charm_from_R}
\end{equation}

\section{The RGE method \label{sec:RGE}}

Another method involves computing $\Delta\hat{\alpha}(M^2_Z)$ first in the $\overline{\mathrm{MS}}$ scheme, following the procedure outlined in Ref.~\cite{Erler:1998sy}.  After that, we can return to the on-shell scheme using \cref{change scheme}. The procedure is:
\begin{itemize}
\item Make use of the nonperturbative input to determine light quark contributions to $\hat{\Pi}(0,\mu^2)$ at a low energy scale ($\mu\approx 2\,\mathrm{GeV}$). If the low energy input is the integral of the experimental $R$ ratio, use \cref{eq:master} and set $s_0=0$. The circle integral in that equation can be evaluated perturbatively using the high-energy expansion of $\hat{\Pi}(s,\mu^2)$ for the light quarks. In contrast, if lattice data is used as an input, the relation \cref{eq:vacuumbothschemes} should be used, with $\Pi(-q^2)$ representing the lattice input. There are two subtleties that need to be addressed during this step: (a) the cross-section data may include virtual corrections from heavy quarks \cite{Hoang:1994it}, and (b) condensates may be of relevance. The two effects are accounted for, resulting in minor corrections.

\item Once the contribution of light quarks has been determined, proceed to apply the matching conditions to add the effects of the charm quark (if one would like to match at the charm mass scale, the RGE has to be used first, see next step). The matching formula, up to order $\hat{\alpha}^2_s$, is provided in \cite{Chetyrkin:1997un}, while at order $\hat{\alpha}^3_s$, it is given in \cite{Sturm:2014nva}. When matching, condensate effects can be incorporated, but their contributions are small.

\item Use  the RGE for $\hat{\alpha}$, known up to order $\hat{\alpha}^4_s$ \cite{Baikov:2012zm}, to run the coupling constant to $\mu=\hat{m}_b$. Apply the matching at that scale and evolve it again with five quarks until $\mu=M_Z$.

\item Finally, convert back to the on-shell scheme using \cref{change scheme}.
\end{itemize}
The following subsections provide a more detailed explanation of each of these steps. Since we will use the high energy expansion of the vacuum polarization function repeatedly, we display it here (it was taken from Refs.~\cite{Nesterenko:2016pmx, Maier:2011jd, Chetyrkin:1996cf, Chetyrkin:2000zk,Shifman:1978bx,Eidelman:1998vc,Surguladze:1990sp,Braaten:1991qm})  
\begin{align}
\label{eq:highenergymassless}
\hat{\Pi}^{(n_q)}(s,\mu^2)^{\mathrm{con}}_{m=0}&=  \sum_i\frac{Q^2_i}{4\pi^2} \Bigg(\frac{5}{3}-l_{s\mu}+  \left( 
\hat{a}_s+\frac{3Q^2_i}{4}\frac{\hat\alpha}{\pi}\right)\left[\frac{55}{12}-4\zeta_3-l_{s\mu}\right]\nonumber \\[12pt]
&+ \hat{a}^2_s \left[\frac{41927}{864}-\frac{829
   \zeta_3}{18}+\frac{25 \zeta_5}{3}+n_q \left(-\frac{3701}{1296}+\frac{19 \zeta_3}{9}\right)\right.\,\nonumber \\[12pt]
&+ \left.\left\{-\frac{365}{24}+11 \zeta_3-n_q \left(-\frac{11}{12}+\frac{2 \zeta_3}{3}\right)\right\}l_{s\mu}+\left\{\frac{11}{8}-\frac{n_q}{12}\right\}l^2_{s\mu}\right]\, \nonumber \\[12pt]
&+ \hat{a}^3_s\left[\frac{31431599}{41472}-\frac{624799\zeta_3}{864}+\frac{1745 \zeta_5}{96}-\frac{665 \zeta_7}{36}+\frac{165 \zeta_3^2}{2}+\frac{11 \zeta_2^2}{24} \right.\, \nonumber \\[12pt]
&+ \left(\frac{545 \zeta_5}{54}-\frac{5 \zeta_3^2}{3}+\frac{174421 \zeta_3}{2592}-\frac{11
   \zeta_2^2}{72}-\frac{1863319}{20736}\right)n_q+\left(\frac{196513}{93312}\right.\, \nonumber \\[12pt]
&- \left.\frac{809 \zeta_3}{648}-\frac{5 \zeta_5}{9}\right)n^2_q+
\left\{-\frac{87029}{288}+\frac{1103 \zeta_3}{4}-\frac{275 \zeta_5}{6}+\left(\frac{7847}{216}\right.\right.\, \nonumber \\[12pt]
&- \left.\left.\frac{262 \zeta_3}{9}+\frac{25 \zeta_5}{9}\right)n_q +\left(\frac{19 \zeta_3}{27}-\frac{151}{162}\right)n_q^2\right\}l_{s\mu}+\left\{\frac{4321}{96}-\frac{121 \zeta_3}{4}\right.\, \nonumber \\[12pt]
&+ \left.\left.\left(\frac{11 \zeta_3}{3}-\frac{785}{144}\right)n_q+\left(\frac{11}{72}-\frac{\zeta_3}{9}\right)n^2_q\right\}l^2_{s\mu}-\frac{1}{48}\left\{11 - \frac{2}{3} 
n_q\right\}^2l^3_{s\mu}\right]\Bigg),\, \nonumber\\[12pt]
& 
\end{align}
\begin{equation}
\hat{\Pi}^{(n_q)}\left(s,\mu^2\right)^{\mathrm{disc}}_{m=0}=\frac{\left(\sum_i Q_i\right)^2}{4\pi^2}\hat{a}^3_s\left[\frac{2155}{2592}-\frac{35
   \zeta_3}{32}+\frac{25 \zeta_5}{24}-\frac{5 \zeta_3^2}{9}-\frac{\zeta_2^2}{12}+\left\{\frac{5 \zeta_3}{9}-\frac{55}{216}\right\} l_{s\mu}\right],\,\,
   \label{eq:highenergyOZI}
\end{equation}
\begin{align}
\hat{\Pi}^{(n_q)}(s,\mu^2)^{\mathrm{con}}_{\mathrm{mass}}&= \sum_{i}\frac{Q^2_i}{4\pi^2}\frac{\hat{m}^2_i}{s}\bigg( 6 +\hat{a}_s\left[16 - 12 l_{s\mu}\right]+\hat{a}^2_s\left[\frac{18923 }{72}+ \frac{196}{9}  \zeta_3  - \frac{1045 \zeta_5}{9}- \frac{95n_q}{12} \right.\nonumber \\[12pt]
&+ \left.\left\{- \frac{253}{2}+ \frac{
 13 n_q }{3} \right\}l_{s\mu}+ \left\{\frac{57  }{2}  - n_q\right\}l_{s\mu}^2\right] \bigg),
\label{eq:highenergymass}
\end{align}
\begin{align}
\Pi^{(3)}_{\mathrm{cond}}(s)&= \sum_{l}Q^2_l\left\{ 
\left(\frac{1}{12}-\frac{11}{216}\hat{a}_s+\hat{a}^2_s\left[c_1+\frac{11}{96}l_{s\mu}\right]\right)\frac{\langle \hat{a}_s G^2\rangle}{s^2}
\right. \nonumber \\[12pt]
&+ 2\left(1+\frac{\hat{a}_s}{3}+\hat{a}^2_s\left[\frac{11}{2}-\frac{3}{4}l_{s\mu}\right]\right)\frac{\langle m\bar{q}q\rangle_l}{s^2} \nonumber \\[12pt]
&\left.\left(\frac{4}{27}\hat{a}_s+\hat{a}^2_s\left[\frac{4 \zeta
   (3)}{3}-\frac{257}{486}-\frac{1}{3}l_{s\mu}\right] \right)\sum_{l^\prime}\frac{\langle m\bar{q} q\rangle_{l^\prime}}{s^2} \right\},
\label{eq:condensateslightmsbar}\end{align}
where $l_{s\mu}\equiv \ln\left(\frac{-s}{\mu^2}\right)$, $n_q$ is the number of quarks, $\hat{m}_q$ is the $\overline{\mathrm{MS}}$ mass of the quark with charge $Q_q$, and  $\zeta_s=\sum^{\infty}_{n=1}n^{-s}$ is the Riemann zeta function. The term $\hat{\Pi}^{(n_q)}(s,\mu^2)^{\mathrm{con}}_{m=0}$ denotes the mass-less and connected contributions to the vacuum polarization function. By connected, we mean that the Feynman diagram cannot be separated if only gluon lines are cut. On the other hand, $\hat{\Pi}^{(n_q)}\left(s,\mu^2\right)^{\mathrm{disc}}_{m=0}$ corresponds to the mass-less disconnected contribution. These are also called OZI suppresed diagrams. 
The term $\hat{\Pi}^{(n_q)}(s,\mu^2)^{\mathrm{con}}_{\mathrm{mass}}$ is the first mass correction to the vacuum polarization. It is obtained through an expansion in terms of the parameter $\frac{\hat{m}^2_i}{s^2}$.
Finally, $\Pi^{(3)}_{\mathrm{cond}}(s)$ represents the three light quark contribution from higher-dimensional operators in the operator product expansion (OPE) \cite{Shifman:1978bx}. The constant $c_1$ will be irrelevant for our purposes.
The vacuum polarization is the sum of the aforementioned contributions.

\subsection{Light quark contribution from low energy $e^+e^-\rightarrow\mathrm{had}$}
The three light quark contribution to $\Delta\hat{\alpha}$ can be computed with the help of \cref{eq:master} with $q^2=0$ and $\mu_0=\mu$, which yields, 
\begin{equation}
    \Delta\hat{\alpha}^{(3)}(\mu^2)=\frac{\alpha}{3\pi}\int^{\mu^2}_{4m^2_\pi}\frac{R(s)}{s}ds-2\alpha i\int_{|s|=\mu^2} \frac{\hat{\Pi}^{(3)}(s,\mu^2)}{s}ds.\label{eq:MS3light}
\end{equation} 
For the first term on the right hand side, we use the low energy  $e^+e^-\rightarrow \mathrm{hadrons}$ integral of DHMZ19 given in \cref{lowintegral}, \begin{equation}
\frac{\alpha M^2_Z}{3\pi}\int^{\mu^2}_{4m^2_\pi}\frac{R(s)}{s(M^2_Z-s+i\epsilon)}ds=\frac{\alpha }{3\pi}\int^{\mu^2}_{4m^2_\pi}\frac{R(s)}{s}ds+\mathcal{O}\left(\frac{\alpha}{3\pi}\frac{\mu^2}{M^2_Z}\right),
\end{equation}
where the term of order $\frac{\alpha}{3\pi}\frac{\mu^2}{M^2_Z}$ is of size $\sim 10^{-7}$ for $\mu\approx\mathrm{2}\,\mathrm{GeV}$ and can be neglected. Since $\mu$ is below the charm threshold, the integral over $R$ in  \cref{eq:MS3light} captures the three light quark contributions. In practice, however, the experimental value of $R$ also incorporates virtual effects from  heavy quarks, which are represented by double-bubble Feynman diagrams with an outer light quark bubble and an inner heavy quark bubble.  Their contribution was previously computed at order  $\hat{a}^2_s$ in Ref.~\cite{Erler:1998sy}. We extend it to order $\hat{a}^3_s$ using the results of Ref.~\cite{Larin:1994va}. The calculation is very similar to the one  in Ref.~\cite{Davier:2023hhn}, where the effects of double bubble diagrams on the Adler function are computed. 

To evaluate the integral over the circle in \cref{eq:MS3light}, we use the large $s$ expansion of the vacuum polarization function in \cref{eq:highenergymassless,eq:highenergyOZI,eq:highenergymass} with $n_q=3$, including the small correction from the strange quark mass. In particular, the disconnected contribution vanishes for three light quarks ($Q_u+Q_d+Q_s=0$).
Using the results \begin{equation}
\int_{|s|=\mu^2}\frac{ds}{s^n}=2\pi i\delta_{1n},\quad\quad\int_{|s|=\mu^2}\frac{l_{s\mu}}{s}ds=0,\quad\quad\int_{|s|=\mu^2}\frac{l^2_{s\mu}}{s}ds=-4i\pi\zeta_2, \label{eq:loginte}
\end{equation}
\begin{equation*}
\int_{|s|=\mu^2}\frac{l^3_{s\mu}}{s}ds=0,\quad\,\int_{|s|=\mu^2}\frac{l_{s\mu}}{s^2}ds=-\frac{2i \pi}{\mu^2}, \quad\,\int_{|s|=\mu^2}\frac{l^2_{s\mu}}{s^2}ds=-\frac{4i\pi}{\mu^2}, \quad\int_{|s|=\mu^2}\frac{l_{s\mu}}{s^3}ds=-\frac{i \pi}{\mu^4} ,
\end{equation*}
we obtain 
\begin{align}
\Delta\hat{\alpha}^{(3)}(\mu^2)&= \frac{\alpha M^2_Z}{3\pi}\int^{\mu^2}_{4m^2_\pi}\frac{R(s)}{s(M^2_Z-s+i\epsilon)}ds+4\pi\alpha\hat{\Pi}^{(3)}(-\mu^2,\mu^2)^{\mathrm{conn}}_{m=0} \nonumber \\[12pt]
&+\frac{2\alpha}{3\pi}\bigg\{{2}\frac{\hat{m}^2_s}{\mu^2} \hat{a}_s+\left[\frac{125}{12}\frac{\hat{m}_{s}^2}{\mu^2}-\frac{9}{4}\zeta_2-A_2\right]\,  \hat{a}^2_s+\biggl[\frac{81}{2}\zeta_2\zeta_3-\frac{961}{16}\zeta_2-A_3\biggl]\hat{a}^3_s \bigg\}\nonumber \\[12pt]
\Delta\hat{\alpha}^{(3)}(4\,\mathrm{GeV}^2)&= \frac{\alpha M^2_Z}{3\pi}\int^{4\,\mathrm{GeV}^2}_{4m^2_\pi}\frac{R(s)}{s(M^2_Z-s+i\epsilon)}ds+\left[24.89-41\,\Delta\hat{\alpha}_s\pm0.18_{\mathrm{tr}}\right]\times10^{-4}.
\label{eq:ms3fromR}
\end{align}
The $\zeta_2$ function commonly arise in the conversion from timelike to spacelike quantities, like in the relationship between the Adler function and $R$. $A_2$ and $A_3$ arise from the subtraction of the double-bubble diagrams referred to previously, and their analytical expressions are provided in the Appendix A.  The contribution of $A_3$ is negligible $\sim \mathcal{O}\left(10^{-7}\right)$. Finally we point out that the  contribution of the vacuum condensates is also negligible. An account of this was briefly provided in Ref.~\cite{Erler:1998sy}, where it was shown that condensates are suppressed by two powers of $\hat{a}_s$,  giving a negligible contribution. Here, we provide a more detailed explanation for this observation. The vacuum condensate contributions arise from the last term on the right-hand side of \cref{eq:MS3light}. Hence,  from \cref{eq:loginte} all the terms give zero contribution except for the logarithmic terms, which after integration lead to $\pi^2$ terms. The result for the integral is then
\begin{equation}
-2\alpha i\int_{|s|=\mu^2} \frac{\Pi^{(3)}_{\mathrm{cond}}(s)}{s}ds=\frac{2\alpha}{3\pi}\hat{a}^2_s\left[\frac{7\pi^2}{6}\frac{\langle m\bar{q} q\rangle_s}{\mu^4}-\frac{11\pi^2}{48}\frac{\langle \hat{a}_s G^2\rangle}{\mu^4}\right]\approx -4\times10^{-8},
\end{equation}
which is completely negligible. Here we have used the value $\langle \hat{a}_s G^2\rangle=0.005\,\mathrm{GeV}^4$ from Ref.~\cite{Dominguez:2014fua} and $\langle m\bar{q} q\rangle_s\approx\,-0.003\,\mathrm{GeV}^4$  from Ref.~\cite{McNeile:2012xh}\footnote{We multiplied the condensate of that reference by $m_s=0.1\,\mathrm{GeV}$.}. Significantly larger values of the gluon condensate are also quoted in the literature, for example Ref.~\cite{Narison:2011xe} quotes $\langle \hat{a}_s G^2\rangle=0.022\pm0.005\,\mathrm{GeV}^4$. Nevertheless, even for such cases, the contribution is still at a negligible level. 

The contribution of the condensates will not be negligible if lattice QCD data is used as input, as will be shown  in the following subsection. Hence, to be on the conservative side, and given the different values in the literature, we will display the parametric dependence on the condensates. For final numerical values we use rather large values with $100\%$ errors
\begin{equation}
\langle \hat{a}_s G^2\rangle=0.01\pm0.01\,\mathrm{GeV}^4\quad\quad\langle m\bar{q} q\rangle_s=\,-0.003\pm0.003\,\mathrm{GeV}^4.
\end{equation}

\subsection{Light quark contribution from lattice\label{LatticeRGE}}

\begin{figure}
     \centering
     \includegraphics[width=0.5\textwidth]{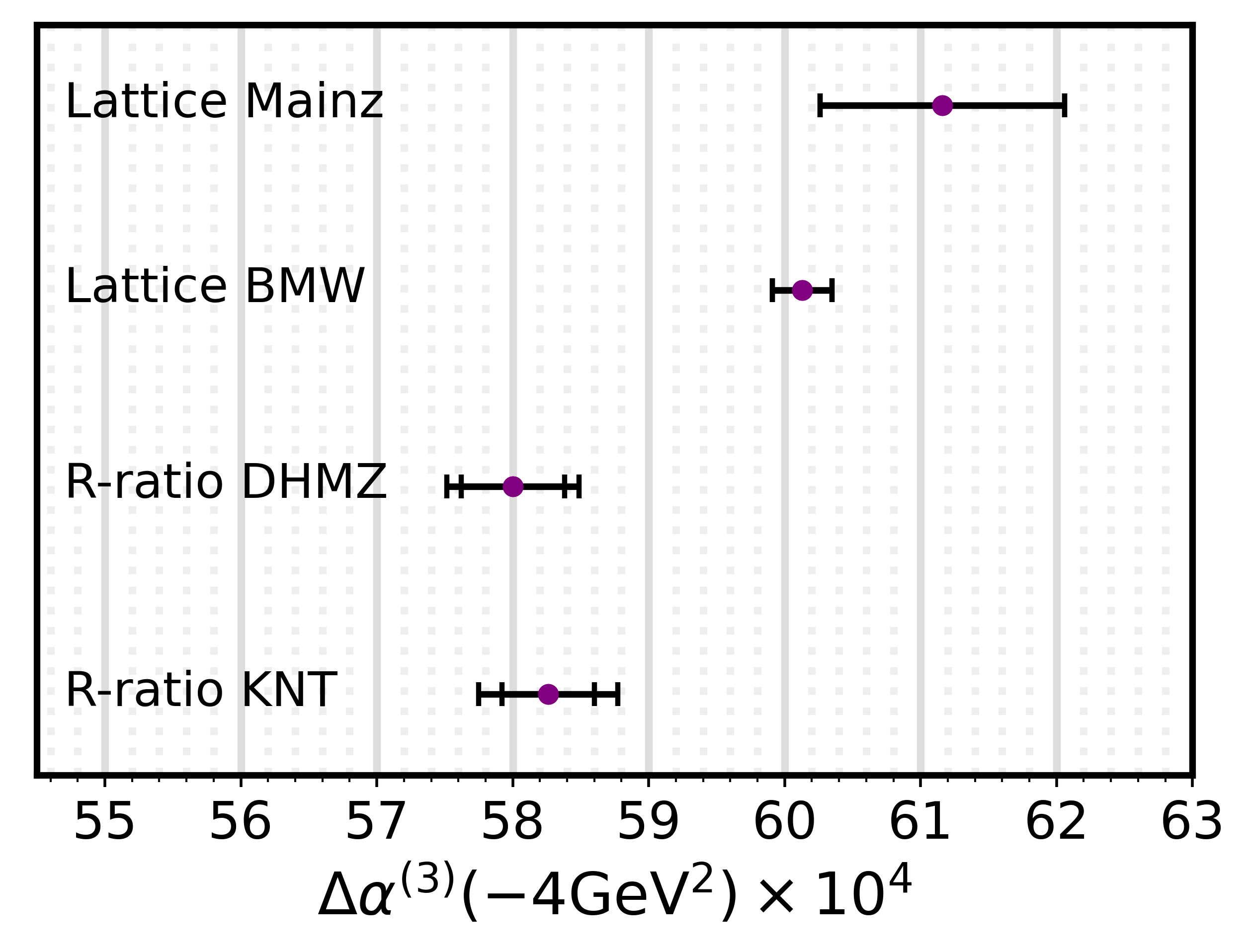}
        
    \caption{Comparison of $\Delta\alpha^{(3)}(-Q^2)$ using \cref{eq:conversionto lattice} to convert the results of Refs.~\cite{Davier:2019can,Keshavarzi:2019abf}. We have used $\sqrt{Q^2}=2\,$GeV, $\hat{\alpha}_s=0.1185\pm0.0016$, and $\hat{m}_s=0.1\,\mathrm{GeV}$. The inner error is the error from data, while the outer includes the error from the conversion.}
    \label{fig:comparison_low3}
\end{figure}

A new contribution of this paper is to use the lattice results from Ref.~\cite{Ce:2022eix} as an input in the RGE method instead of the $R$ ratio data.  From \cref{change scheme} and \cref{eq:highenergymassless,eq:highenergymass,eq:condensateslightmsbar} it is possible to see that $\left(Q^2=-q^2>0\right)$ \begin{align}
\Delta\hat{\alpha}^{(3)}(\mu^2)&= \Delta \alpha^{(3)}(-Q^2)+4\pi\alpha \,\hat{\Pi}^{(3)}(-Q^2,\mu^2)\label{eq:threemslattice}
 \\
&=\Delta \alpha^{(3)}(-Q^2)+\left[25.61\pm0.07_{\mathrm{tr}}+4\,\Delta\hat{\alpha}_s+\frac{48\langle \hat{a}_sG^2\rangle}{Q^4}+\frac{219}{Q^4}\langle m\bar{q} q\rangle_s\right]\times10^{-4}
\nonumber 
\end{align}
where we have set $\mu^2=Q^2=\left(2\,\mathrm{GeV}\right)^2$. The Mainz collaboration \cite{Ce:2022eix} quotes
\begin{equation}
\Delta \alpha^{(3)}(-4\,\mathrm{GeV}^2)=\left(61.2\pm0.9\right)\times10^{-4}\,\,\,\, \,\,\,(\mathrm{Mainz\,Collaboration}),\label{latticelow}
\end{equation}
 while using the running from Ref. \cite{Budapest-Marseille-Wuppertal:2017okr} and with the help of the supplementary section of Ref. \cite{Borsanyi:2020mff} we estimate the BMW collaboration  result\footnote{For definiteness and notwithstanding the preliminary status pointed out in Ref. \cite{Davier:2023cyp}  we use the results of Ref. \cite{Borsanyi:2020mff} as quoted.
Any future update will be straightforward to apply.}  at the same scale to be\footnote{The reader should note that Ref. \cite{Borsanyi:2020mff} gives only $\Pi(-1\,\mathrm{GeV}^2)$ and  $\Pi(-1\,\mathrm{GeV}^2)-\Pi(-10\,\mathrm{GeV}^2)$. On the other hand,  Ref. \cite{Budapest-Marseille-Wuppertal:2017okr} gives $\Pi$ at different points but with larger errors. Hence, to compute the central value at $4\,\mathrm{GeV}^2$ we took the difference between 
 $\Pi(-1\,\mathrm{GeV}^2)$ and $\Pi(-4\,\mathrm{GeV}^2)$ from Ref. \cite{Budapest-Marseille-Wuppertal:2017okr} and added this to the result of  Ref.  \cite{Borsanyi:2020mff}. As a cross check, we added $\Pi(-1\,\mathrm{GeV}^2)$ and $\Pi(-1\,\mathrm{GeV}^2)-\Pi(-10\,\mathrm{GeV}^2)$ from Ref.  \cite{Borsanyi:2020mff} to get $\Pi(-10\,\mathrm{GeV}^2)$ and then ran this down to $4\,\mathrm{GeV}^2$ using perturbative QCD, the results were in agreement to each other. This also confirms that lattice and pQCD agree in the high energy limit. Finally we also assumed that the error at  $4\,\mathrm{GeV}^2$ would be roughly the same as the one at  $1\,\mathrm{GeV}^2$.    } 
 \begin{equation}
\Delta \alpha^{(3)}(-4\,\mathrm{GeV}^2)=\left(60.13\pm0.22\right)\times10^{-4}\,\,\,\, \,\,\,(\mathrm{BMW\,Collaboration}).\label{latticelowBMW}
\end{equation}
 To conclude this subsection we would like to note that it is possible now to relate the timelike integral given in \cref{lowintegral}  to the Euclidean result in \cref{latticelow,latticelowBMW}, to do so we equate the first line of \cref{eq:threemslattice}  to \cref{eq:ms3fromR},\begin{align}
  \Delta\alpha^{(3)}(-Q^2)
 \,\, &=\frac{\alpha M^2_Z}{3\pi}\int^{Q^2}_{4 m^2_\pi}\frac{R(s)}{s(s-M^2_Z)}ds+\frac{2\alpha}{3\pi}\left\{\frac{\hat{m}^2_s}{Q^2}+\hat{a}_s\frac{14}{3}\frac{\hat{m}^2_s}{Q^2}\right.
  \nonumber \\
  &+\hat{a}_s^2\left[\frac{\hat{m}^2_s}{Q^2}\left(\frac{98 \zeta_3}{27}-\frac{1045 \zeta (5)}{54}+\frac{21713}{432}\right)-\frac{9}{4}\zeta_2-A_2\right]\label{eq:conversionto lattice}
 \\
     &+\left.\hat{a}_s^3\left[\frac{81}{2}\zeta_2\zeta_3-\frac{961}{16}\zeta_2\right]\right\}-\left[\frac{48\langle \hat{a}_sG^2\rangle}{Q^4}+\frac{219}{Q^4}\langle m\bar{q} q\rangle_s\right]\times10^{-4},\nonumber
\end{align}
which allows for a direct comparison between lattice  QCD data and $R$ ratio experimental results as shown in \cref{fig:comparison_low3} where we have also included the results from Ref.~\cite{Keshavarzi:2019abf}.
This is expression is highly convenient, since it involves only information from light quarks. 

\subsection{RGE}
The vacuum polarization function satisfies the renormalization group evolution equation \begin{equation}
4\pi^2\mu^2\frac{d}{d\mu^2}\hat{\Pi}(s,\mu^2)=\beta\quad\quad\quad \mu^2\frac{d\hat{a}}{d\mu^2}=\hat{a}^2\beta,
\end{equation}
where $\beta$ is the beta function of the $\overline{\mathrm{MS}}$ coupling  $\hat{a}\equiv \hat{\alpha}/\pi$. In terms of $\Delta\hat{\alpha}$ this may be written as 
\begin{equation}
\mu^2\frac{d}{d\mu^2}\Delta\hat{\alpha}=\frac{\alpha}{\pi}\beta,\label{eq:RGEalpha}
\end{equation}
 The beta function was obtained at 5 loops in Ref.~\cite{Baikov:2012zm}.
The contribution from the quarks is given by
\begin{equation}
\beta= \left[\frac{1}{3} \sum_q K_q  Q_q^2 + 
\sigma\left(\sum_q Q_q \right)^2 \right],
\label{eq:fine structure}
\end{equation}
where $K_q$ represents contributions from connected bubble diagrams that contain gluon insertions in the quark bubble and, as a consequence, can be expanded as a power series in $
\hat{\alpha}_s$. $\sigma$ represents OZI rule suppressed terms, that is, diagrams which become disconnected upon cutting gluon lines, and which only start to contribute at order $
\hat{\alpha}^3_s$. We refer the reader to the Appendix B for a definition of each term. The  RGE can either be solved numerically, or expressing the solution using the QCD beta function and $\hat{\alpha}_s$ at two scales, which  is convenient for global fits given that $\hat{\alpha}_s$ is already computed in available codes \cite{Herren:2017osy,Erler:2011ux}. Appendix B gives that formula. For a quick estimate of the truncation error, we note that the $\hat{a}^4_s$ term in the beta function is approximately  $\approx-10.\,\hat{a}^{4}_s$ for 5 active quarks. Assuming an average $\hat{\alpha}_s\approx 0.2$ within the $2-91\,\mathrm{GeV}$ range, we expect the contribution to the running from the last term in the RGE to be $\sim -3\times10^{-6}$, which agrees with the exact calculation.  The mixed $\hat{\alpha}\hat{\alpha}_s$ term (not shown) is negligible, with an expected contribution of $\sim10^{-7}$~\cite{Erler:1998sy}.

\subsection{Matching}

In the effective field theory framework there is a matching procedure \cite{PhysRevD.11.2856}, meaning that $\hat{\alpha}$ is rescaled  as $\hat{\alpha}^{(n_q-1)}(\mu)=\zeta^2_{\gamma}(\mu) \hat{\alpha}^{(n_q)}(\mu)$, when the renormalization scale $\mu$ is below a particle's decoupling scale. The decoupling constant\footnote{ Refer to \cite{Grozin:2005yg} for an introductory explanation.} $\zeta^2_{\gamma}(\mu)$ has been computed  at order $\hat{a}^2_s$ \cite{Chetyrkin:1997un} and order $\hat{a}^3_s$ \cite{Sturm:2014nva}. For the sake of readability and implementation, we provide here numerical expressions for the  decoupling with $n_q=4$ and $n_q=5$ (charm and bottom quarks):\begin{equation}
\Delta\hat{\alpha}^{(3)}(\hat{m}^2_c)=\Delta\hat{\alpha}^{(4)}(\hat{m}^2_c)-3.0\times10^{-6}-0.001118\,  \hat{a}_s-0.002551\,  \hat{a}^{2}_s+0.00131\, \hat{a}^{3}_s, \label{eq:charmmatchnumeric}
\end{equation}
\begin{equation}
\Delta\hat{\alpha}^{(4)}(\hat{m}^2_b)=\Delta\hat{\alpha}^{(5)}(\hat{m}^2_b)-2\times10^{-7}-0.000280\, \hat{a}_s-0.001209\,\hat{a}_s^2-0.00158\,\hat{a}_s^3.\label{eq:bottommmatchnumeric}
\end{equation}
In Appendix D, the full analytical expression is presented including the effects of disconnected and double bubble diagrams. The size of the $\hat{a}^3_s$ term is $2.6\times10^{-6}$ and $-0.6\times10^{-6}$ for the charm and bottom quarks respectively, and is taken as the perturbative error of the matching.  
The condensates also contribute to the vacuum polarization function of the heavy quarks \cite{Broadhurst:1994qj} and, consequently, to the matching. Specifically, for the charm quark \begin{equation}
    4\pi\alpha\hat{\Pi}^{(c)}_{\mathrm{cond}}(0)=\pi\alpha Q^2_c\left[-\frac{1}{30}\left(1+\frac{605}{162}\hat{a}_s(\hat{m}_c)\right)\frac{\langle\hat{a}_s G^2\rangle}{\hat{m}^4_c}\right]=-\frac{1.8}{\mathrm{GeV}^4}\langle\hat{a}_s G^2\rangle\times10^{-4},\label{eq:charmcondensatesmatching}
\end{equation}
while the contribution of the condensates to the bottom quark matching is negligible. It is worth noting that although small, the contribution from the condensate to the charm matching is comparable in magnitude to the perturbative error from the matching and should be included.

 \subsection{Returning to the on-shell scheme}
 The final step is then to convert from the $\overline{\mathrm{MS}}$ scheme to the on-shell scheme, see \cref{change scheme}.
We use the high energy expansion of the vacuum polarization function \cref{eq:highenergymassless,eq:highenergymass,eq:highenergyOZI} with $n_q=5$, namely  \begin{align}
 \Delta\alpha^{(5)}(M^2_Z) &   = \Delta\widehat\alpha^{(5)}(M^2_Z)-\frac{\alpha}{\pi} \Bigg\{\frac{55}{27}
+\frac{2}{3}\frac{\hat{m}^2_b(M_Z)}{M^2_Z} +\hat{a}_s(M_Z) \left[ \frac{605}{108} - \frac{44}{9} \zeta_3 \right] 
\cr\noalign{\vskip 4pt}
&+ \hat{a}^2_s(M_Z) \left[ \frac{976481}{23328} - \frac{253}{36} \zeta_2 - \frac{781}{18} \zeta_3 + \frac{275}{27} \zeta_5 \right]\cr\noalign{\vskip 4pt}
&+  \hat{a}^3_s(M_Z)\left[\frac{1483517111}{3359232}-\frac{22781 \zeta_2}{144}-\frac{3972649 \zeta_3}{7776}-\frac{31
  \zeta_2^2}{81}+\frac{521255 \zeta
   (5)}{7776}\right.\cr\noalign{\vskip 4pt}
&-  \left. \frac{7315 \zeta (7)}{324} +\frac{5819 \zeta_3 \zeta_2}{54}+\frac{14675 \zeta_3^2}{162}\right]  \Bigg\}.
\label{eq:alphaconversion}
\end{align}
Using the $R$ ratio low energy data in  \cref{lowintegral,eq:ms3fromR} as input, adding the running  given by \cref{eq:RGEalpha}, the matching given in \cref{eq:charmmatchnumeric,eq:bottommmatchnumeric} and the conversion to the on-shell scheme in \cref{eq:alphaconversion}, we obtain \begin{align}
  \Delta\alpha^{(5)}(M^2_Z)
  =&\frac{\alpha M^2_Z}{3\pi}\int^{\mu^2}_{4m^2_\pi}\frac{R(s)ds}{s(M^2_Z-s)}+\bigg[ 217.60-16.9\,\ln\frac{\mu^2}{4\,\mathrm{GeV}^2}+306\,\Delta\hat{\alpha}_s\,\,\,\,
  \notag\\
     -&20\Delta \hat{m}_c-1.3\,\Delta\hat{m}_b-\frac{1.8}{\mathrm{GeV}^4}\langle\hat{a}_s G^2\rangle\pm 0.20_{\mathrm{tr}}\bigg]\times10^{-4} \notag\\
     =&\left(276.29\pm0.73\right)\times10^{-4}.\label{eq:RGERratio}
\end{align}
 On the other hand, using lattice QCD results as input data with the help of  \cref{eq:threemslattice}, adding the running  given by \cref{eq:RGEalpha}, the matching  given in \cref{eq:charmmatchnumeric,eq:bottommmatchnumeric} and the conversion to the on-shell scheme given in \cref{eq:alphaconversion}, we obtain 
\begin{align}
  \Delta\alpha^{(5)}(M^2_Z)
  =&\Delta\alpha^{(3)}(-Q^2)+\bigg[ 218.32-17.3\,\ln\frac{Q^2}{4\,\mathrm{GeV}^2}+350\,\Delta\hat{\alpha}_s-20\Delta \hat{m}_c\,\,\,\,
     \notag\\
     &-1.3\,\Delta\hat{m}_b-\left(\frac{1.8}{\mathrm{GeV}^4}-\frac{48}{Q^4}\right)\langle \hat{a}_sG^2\rangle+\frac{219}{Q^4}\langle m\bar{q} q\rangle_s \pm 0.05_{\mathrm{tr}}\bigg]\times10^{-4} \notag\\
     =&\left(279.5\pm1.1\right)\times10^{-4}\quad\quad\,\,\,\,\,\,\quad (\mathrm{Mainz\,Collaboration})\notag\\
     =&\left(278.42\pm0.63\right)\times10^{-4}\quad\quad\quad (\mathrm{BMW\,Collaboration}).\label{eq:RGElattice}
\end{align}
 The charm quark contribution to these expressions is given by 
\begin{equation}
  \Delta\alpha^{(c)}(M^2_Z)
  =\left[79.47-\frac{1.8}{\mathrm{GeV}^4}\langle\hat{a}_s G^2\rangle+207\,\Delta\hat{\alpha}_s
     -20\,\Delta \hat{m}_c\pm0.02_{\mathrm{tr}}\right]\times 10^{-4}.
     \label{MScharm}
\end{equation}
This result is in agreement with the calculation done in Ref.~\cite{Dominguez:2014fua} where this contribution was estimated using pQCD and sum-rules.
A summary of our results can be found in \cref{tab:MSBAR}.
\begin{table}[h]
\begin{center}
\begin{tabular}{ |p{1.2cm}||p{2.1cm}|p{2.3cm}|p{2.1cm}| p{2.1cm}||  p{2.1cm}|  }

 \hline
 \multicolumn{6}{|c|}{Contributions to $\Delta\alpha^{(5)}(M_Z)\times10^{-4}$ RGE method R input by type} \\
 \hline
 Source& Non-pert (low energy and condensates)  & Matching (pert) & Running & Conversion & Total\\
 \hline
u, d, s  & 58.71$\pm$\textcolor{blue}{0.38}$\pm$ \textcolor{Thistle}{0.28}    &24.89$\pm$\textcolor{OliveGreen}{0.18}$\pm$ \textcolor{BurntOrange}{0.07} & 125.87$\pm$\textcolor{OliveGreen}{0.04}$\pm$ \textcolor{BurntOrange}{0.19}&-25.44$\pm$ \textcolor{OliveGreen}{0.03}$\pm$ \textcolor{BurntOrange}{0.01} &184.02$\pm$\textcolor{OliveGreen}{0.18}$\pm$ \textcolor{BurntOrange}{0.13}$\pm$\textcolor{blue}{0.38}$\pm$ \textcolor{Thistle}{0.28}\\
\hline
 charm &   -0.02 $\pm$ 0.02  & 1.81$\pm$\textcolor{OliveGreen}{0.03} $\pm$ \textcolor{BurntOrange}{0.12}$\pm$ \textcolor{Mahogany}{0.01}   &94.62 $\pm$\textcolor{OliveGreen}{0.01} $\pm$\textcolor{BurntOrange}{0.20}$\pm$ \textcolor{Mahogany}{0.15} &-16.96$\pm$\textcolor{OliveGreen}{0.02} $\pm$ \textcolor{BurntOrange}{0.01} &79.45$\pm$\textcolor{OliveGreen}{0.02} $\pm$\textcolor{BurntOrange}{0.33}$\pm$ \textcolor{Mahogany}{0.16}$\pm$0.02  \\
 \hline
  bottom & 0.00 & 0.27$\pm$\textcolor{OliveGreen}{0.01} $\pm$\textcolor{BurntOrange}{0.01}&  16.80 $\pm$\textcolor{BurntOrange}{0.02}  $\pm$0.01$_{m_b}$&-4.26&12.82$\pm$\textcolor{OliveGreen}{0.01} $\pm$\textcolor{BurntOrange}{0.03}$\pm$0.01$_{m_b}$\\
 \hline
 OZI   & 0.00 & 0.00& -0.02$\pm$\textcolor{OliveGreen}{0.01} &0.00 &-0.02$\pm$\textcolor{OliveGreen}{0.01}\\
 \hline
 QED$^*$  &   0.00  & 0.03 & 0.14& 0.00 &0.18\\
 \hline
 \hline
 Total  & 58.69$\pm$\textcolor{blue}{0.38}$\pm$ \textcolor{Thistle}{0.28}$\pm$0.02  & 26.97 $\pm$\textcolor{OliveGreen}{0.20}$\pm$\textcolor{BurntOrange}{0.06}$\pm$ \textcolor{Mahogany}{0.01}   &237.29$\pm$\textcolor{OliveGreen}{0.06} $\pm$\textcolor{BurntOrange}{0.42}$\pm$ \textcolor{Mahogany}{0.15}$\pm$0.01$_{m_b}$ &-46.66$\pm$\textcolor{OliveGreen}{0.05} $\pm$\textcolor{BurntOrange}{0.02}&276.29$\pm$\textcolor{OliveGreen}{0.20} $\pm$\textcolor{BurntOrange}{0.49} $\pm$\textcolor{Mahogany}{0.16}$\pm$\textcolor{blue}{0.38}$\pm$ \textcolor{Thistle}{0.28}$\pm$0.02 $\pm$0.01$_{m_b}$\\
\hline
\end{tabular}
\end{center}
\caption{\label{tab:MSBAR} Contributions to $
\Delta\alpha^{(5)}(M^2_Z)$.  Orange: parametric error from $\hat{\alpha}_s$ using  $\hat{\alpha}_s(M_Z)=0.1185\pm0.0016$. Green: perturbative error defined as the size of the last known term in the $\hat{\alpha}_s$ expansion. Blue: error from the low energy data. Pink: Quark hadron duality error. Brown: parametric error coming from the charm mass using $\ \hat{m}_c(\hat{m}_c)=1.270\pm0.008\,\mathrm{GeV}$. For the bottom quark we use as reference $\ \hat{m}_b(\hat{m}_b)=4.180\pm0.008\,\mathrm{GeV}$. OZI denotes the disconnected diagrams, which give a very small contribution. If the vacuum polarization function in the spacelike is used as  input, there will be a change in the u,d,s- Non-pert box and u,d,s-Matching box. The u,d,s-Matching box changes to $25.61\pm\textcolor{OliveGreen}{0.07}\pm\textcolor{BurntOrange}{0.01}$ from  \cref{eq:threemslattice} (the condensate contribution should be included too),  while the  u,d,s- Non-pert box changes to  $61.2\pm\textcolor{blue}{0.9}$ or $60.13\pm\textcolor{blue}{0.22}$  for $\mu=2\,\mathrm{GeV}$ given in \cref{latticelow,latticelowBMW} respectively.   *The QED row contains the two loop QED effects, with terms proportional to $Q_q^4$, and  should not be added in this table since it is already included in the contribution of each quark. }
\end{table}
\section{Euclidean split technique}

The Euclidean split technique was introduced in Refs.~\cite{Jegerlehner:2008rs,Proceedings:2019vxr}. It makes use of the Adler function, which is defined as 
\begin{equation}
    D(Q^2)\equiv-12 \pi^2 Q^2\frac{d\Pi(-Q^2)}{dQ^2}=3\pi\frac{d\Delta\alpha(-Q^2)}{dQ^2}.\label{Adler}
\end{equation}
In the high energy limit the Adler function can be computed as a power series in $\hat{\alpha}_s$.  It can also be computed as an integral over data, by taking the derivative of \cref{eq:onshellintegral}, which leads to 
\begin{equation}
D(Q^2)=Q^2\int^{\infty}_{4m^2_\pi}\frac{R(s)}{(s+Q^2)^2}ds.\label{Adlerdata}
\end{equation}
By comparing the perturbative and data driven values of $D(Q^2)$, one can find $Q^2_0$, which is defined as the minimal scale where both, data and pQCD agree with each other. Its value is around $\sim 2\,\mathrm{GeV}$ as shown in Ref.~\cite{Jegerlehner:2008rs}. The idea of the Euclidean split technique is to use $Q^2_0$ as a leverage to split $\Delta\alpha(M^2_Z)$ into several contributions 
\begin{equation}
    \Delta\alpha(M^2_Z)= \Delta\alpha(-Q^2_0)+ \left[\Delta\alpha(-M^2_Z)- \Delta\alpha(-Q^2_0)\right]_{\mathrm{pQCD}}+\left[\Delta\alpha(M^2_Z)- \Delta\alpha(-M^2_Z)\right]_{\mathrm{pQCD}}.
    \label{eq:eucspliting}
\end{equation}
Note that here just zeros have been added to $\Delta\alpha(M^2_Z)$, it may look as a trivial relation. Nevertheless, this detour allows to have a good control of the error. In Ref.~\cite{Proceedings:2019vxr} The first term  in \cref{eq:eucspliting} is computed using \cref{eq:onshellintegral}
    \begin{equation}
    \Delta\alpha(-Q^2_0)_{\mathrm{dat+pQCD}}=\frac{\alpha Q^2_0}{3\pi}\int^{\infty}_{4m^2_\pi}\frac{R(s)}{s(s+Q^2_0)}ds,
    \label{eq:onshellintegraleuclidean}
\end{equation} 
where the integral can be divided into a region where data is used and another where pQCD is applicable. Furthermore, given the low value of $Q_0^2$, the kernel makes this integral to be dominated by the low energy data. The left hand side of \cref{eq:onshellintegraleuclidean} can also be obtained directly from lattice QCD, as shown in Ref.~\cite{Ce:2022eix}.  The second term in \cref{eq:eucspliting} gives the change from the low energy scale $Q^2_0$ to $M^2_Z$ and can be computed using the perturbative expansion of the Adler function. The last term takes us back to the timelike. It can easily be 
 computed within pQCD  with the help of the high energy expansion of the vacuum polarization function for five quarks given in \cref{eq:highenergymassless}, which gives
\begin{equation}
    \left[\Delta\alpha(M^2_Z)-\Delta\alpha(-M^2_Z)\right]_{\mathrm{pQCD}}=\frac{\alpha}{\pi}  \left(-\frac{4\hat{m}^2_b}{3M^2_Z}{}+\hat{a}_s^2\frac{253
   \zeta_2}{36}+\hat{a}_s^3 \left[\frac{22781 \zeta_2}{144}-\frac{5819 \zeta_3 \zeta_2}{54}\right]\right),
    \label{eq:onshelltoeuclideanMz}
\end{equation} 
where $\hat{a}_s$ is evaluated at $\mu=M_Z$ for 5 active quarks. 

In our calculations, we used the $\overline{\mathrm{MS}}$  renormalisation scheme for $\hat{\alpha}_s$ instead of the momentum substraction scheme (MOM) \cite{Jegerlehner:1998zg,Eidelman:1998vc}. Furthermore, instead of performing  the full  integral shown in \cref{eq:onshellintegraleuclidean},  we use the result of the already computed integral in DHMZ19 and transform it with the help of \cref{eq:conversionto lattice} adding the charm and bottom contributions with the heavy quark pQCD expansion of $\Pi$.

To compute the integral of the pQCD Adler function we interpolated\footnote{We used a spline interpolation, but the final result changes marginally when another interpolation is done. Differences are included in the truncation error. A more detailed analysis focused on the Adler function itself and its scheme dependence is under preparation.  } its low and high  energy expansions. 
Our result is\begin{align}
  \Delta\alpha^{(5)}(M^2_Z)
  =&\left[276.42+316\,\Delta\hat{\alpha}_s- 20\,\Delta\hat{m}_c-1.3\,\Delta\hat{m}_b\pm 0.30_{\mathrm{tr}}\pm0.38_{\mathrm{data}<2\,\mathrm{GeV}}\right.
  \nonumber\\
    &\pm\left.0.28_{\mathrm{dual}}\right]\times 10^{-4}
    \label{Adler5E}
\end{align}
 It is quite reassuring to see the excellent agreement between this and the RGE method \cref{eq:RGERratio} since both use the same input data. This result is also in agreement with the integration in the timelike given in \cref{eq:ExplicitIntegration}. We remark that for the contribution of the charm quark, we also have an excellent agreement with the RGE method in \cref{MScharm}.

\section{Discusion}

\begin{figure}
     \centering
     \begin{subfigure}[b]{0.45\textwidth}
         \centering
         \includegraphics[width=\textwidth]{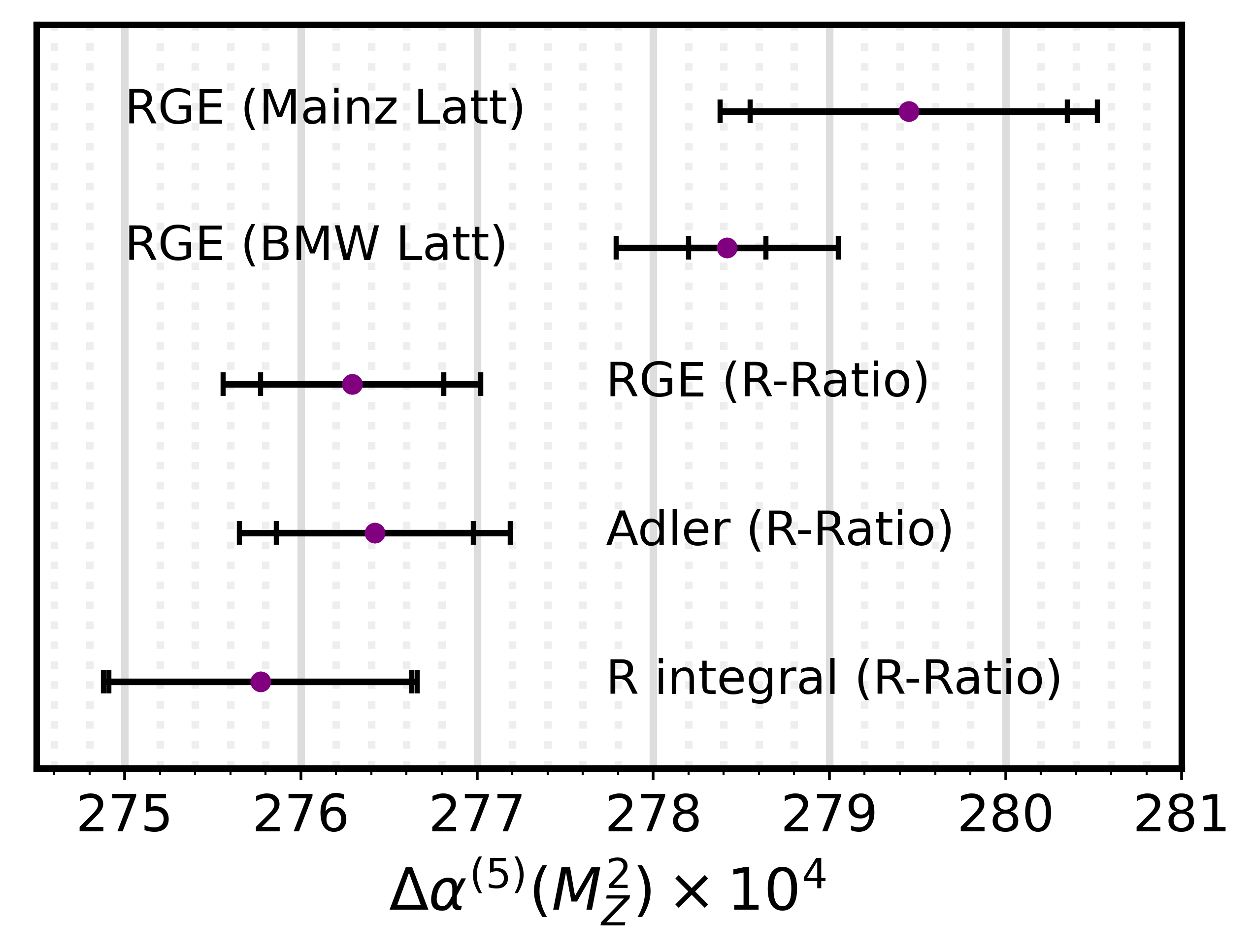}
        
     \end{subfigure}
     \begin{subfigure}[b]{0.45\textwidth}
         \centering
         \includegraphics[width=\textwidth]{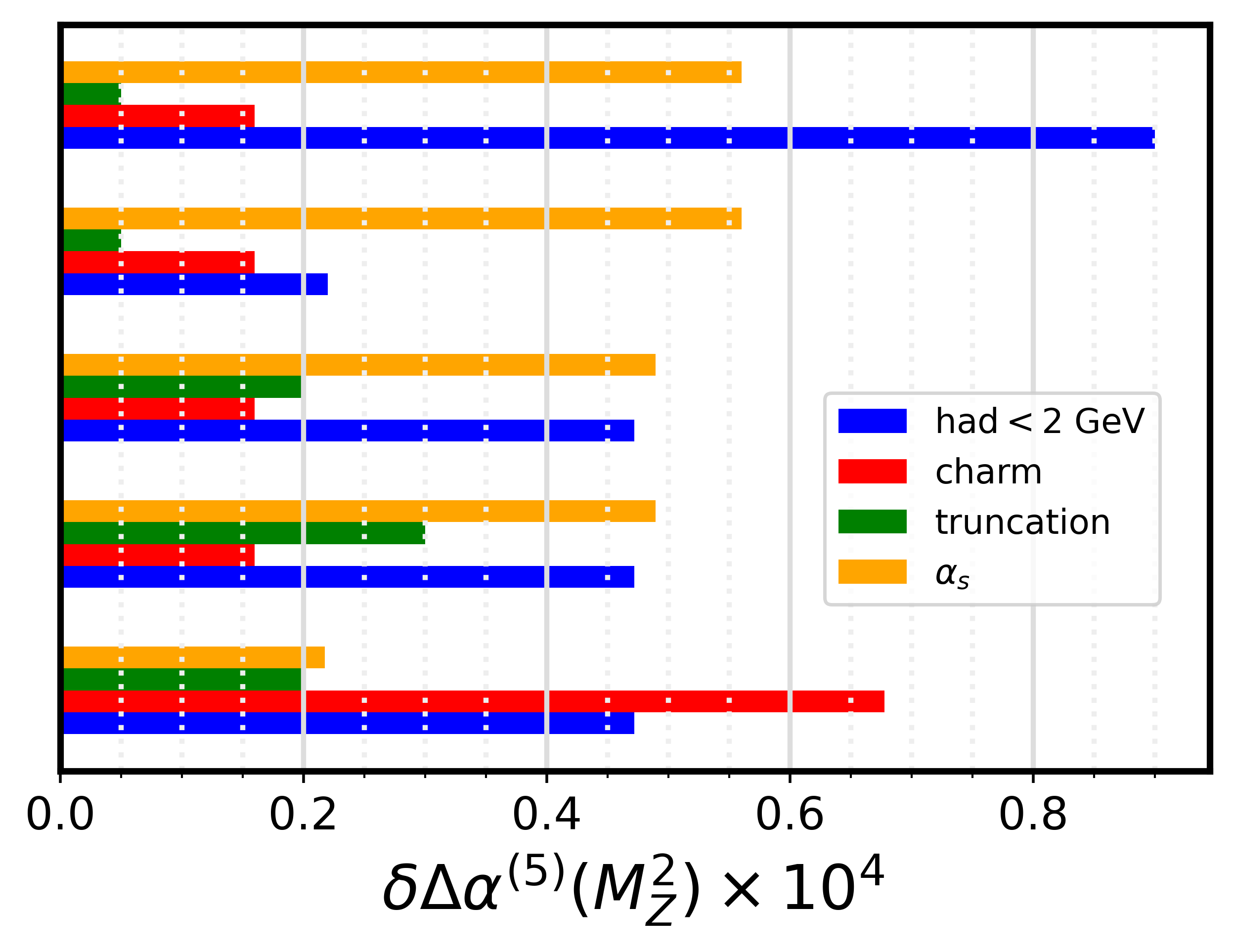}
        
     \end{subfigure}
     \hfill
        \caption{Left: $\Delta\alpha^{(5)}(M_Z)$ in the different frameworks. The inner error corresponds to the non parametric error. Right: individual error sources.  }
        \label{fig:Dalpha5all}
\end{figure}

The results presented in \cref{eq:ExplicitIntegration,eq:RGERratio,Adler5E} indicate that all three methods using $R$ data as input yield consistent results within the non-parametric error. As can be seen in \cref{fig:Dalpha5all}, \cref{eq:ExplicitIntegration} exhibits greater error and smaller central value, where the significant difference originates primarily from the charm quark sector. 

In the RGE approach, the error associated with the charm quark stems from the uncertainty in the input parameter $\hat{m}_c$. However, in the explicit integration method, the charm quark error can be attributed to several sources, including resonances, the continuum (with the highest impact coming from the integral in the range $[3.7,5]\,\mathrm{GeV}$), and to a lesser extent, the input parameter $\hat{m}_c$.
\begin{figure}
     \centering
         \includegraphics[scale=0.7]{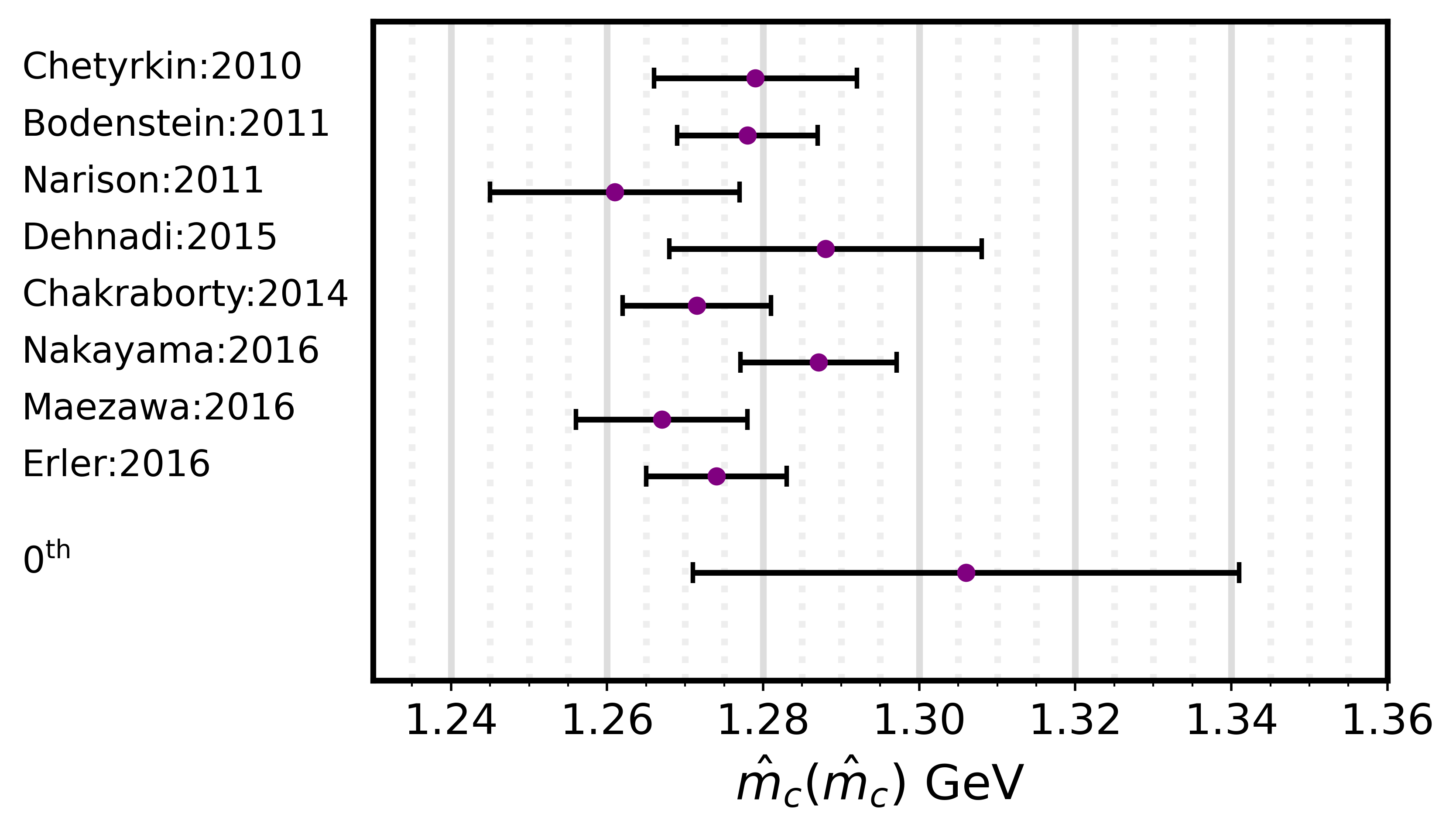}
         \caption{Charm quark mass from different collaborations \cite{Chetyrkin:2010ic, Bodenstein:2011ma, Narison:2011xe, Dehnadi:2015fra, Chakraborty:2014aca,Nakayama:2016atf, Maezawa:2016vgv, Erler:2016atg}. The last value corresponds to computing the charm mass from the zeroth moment, taking an error of $\delta\alpha_s=0.0016$. }
        \label{fig:charmmass}
\end{figure}
We can relate the charm contribution to $\Delta\alpha^{(c)}(M^2_Z)$ in the explicit integration method to a value of the charm quark mass. To accomplish this, we equate \cref{eq:charm_from_R} and \cref{MScharm} to each other and solve for $\hat{m}_c$. The result is 
\begin{equation}
   \hat{m}_c=1.305\,\mathrm{GeV} +8.6(\Delta\hat{\alpha}_s)\pm0.032_{\mathrm{c-thr}}\pm0.007_{J/\psi}\pm0.001_{\mathrm{tr}},\label{charmmasscompared}
\end{equation}
which is in agreement with other determinations but has a larger error, see \cref{fig:charmmass}. This result can be understood as an independent extraction of the charm quark mass using only the zeroth moment \cite{Erler:2016atg}.  

In principle one may argue that the same data is used to extract the charm mass as in Refs.~\cite{Erler:2016atg} or \cite{Kuhn:2007vp}, but the relevance of data is different as we explain now.  The charm quark mass is usually extracted from heavy quark moments, which are defined as the coefficients in a Taylor expansion around $q^2=0$ of the vacuum polarization function of a heavy quark. For $n>0$ the moments are defined as 
\begin{equation}
    \mathcal{M}_n\equiv \frac{12\pi^2}{n!}\left.\frac{d^n}{dt^n} \hat{\Pi}_q(t)\right|_{t=0}=\mathcal{M}^{\mathrm{pQCD}}_n+\mathcal{M}^{\mathrm{cond}}_n,\label{momentsdef}
\end{equation}
where the perturbative contributions are known up to $\mathcal{O}\left(\hat{\alpha}^4_s\right)$ \cite{Chetyrkin:2006xg}
\begin{equation}
    \mathcal{M}^{\mathrm{pQCD}}_n=\frac{9}{4}Q^2_q \left(\frac{1}{2\hat{m}_q(\hat{m}_q)}\right)^{2n}\hat{C}_n \quad\quad\quad\quad\quad  \hat{C}_n=\sum^3_{k=0}{\hat{a}_s^{k}C^{(k)}_n}+\mathcal{O}\left(\hat{\alpha}^4_s\right),\label{momentspert}
\end{equation}
while the contribution from the condensates is \cite{Broadhurst:1994qj}  \begin{equation}
    \mathcal{M}^{\mathrm{cond}}_n=\frac{12\pi^2Q^2_q}{\left(4\hat{m}^2_q\right)^{n+2}}\left<\hat{a}_s G^2\right>a_n\left(1+\hat{a}_s b_n\right).
\end{equation}$a_n$ and $b_n$ are constants whose value is given in Ref. \cite{Broadhurst:1994qj}. To extract the quark mass, one has to relate these moments to an experimental measurement. By computing the derivatives of \cref{eq:pifromR} it is straightforward to see that 
\begin{equation}
\mathcal{M}_n=\int^{\infty}_{4\hat{m}^2_q}\frac{R_q(s) ds}{s^{n+1}},\label{momentsintegral}
\end{equation}
where $R_q$ is the contribution to $R$ from a specific quark $q$.

\begin{figure}
     \centering
         \includegraphics[scale=0.6]{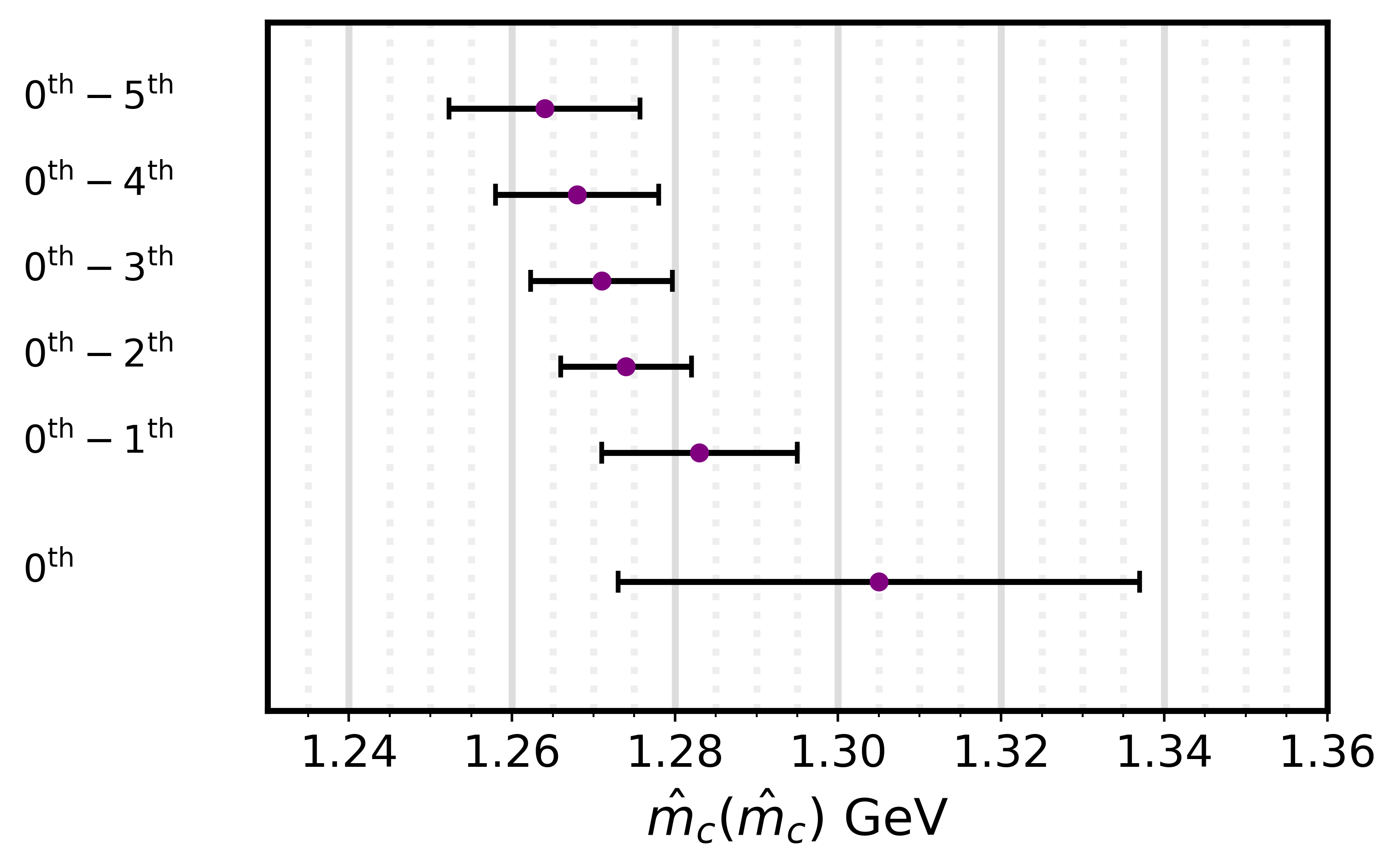}
         \caption{Extraction of the charm mass from combinations of different moments from Ref. \cite{Erler:2016atg}, the last line corresponding to the extraction from the $0^{\mathrm{th}}$ moment is the result of this work. One can see that the tendency for higher values of the mass as lower moments are used continues. Furthermore, in the extraction form the $0^{\mathrm{th}}-1^{\mathrm{th}}$ combination it was already seen that the dominant part of the error came from the charm threshold region. }
        \label{fig:moments}
\end{figure}

A determination of the charm mass can be made by comparing the experimental moment value, obtained from \cref{momentsintegral}, with the theoretical expression (the right side of \cref{momentsdef}). Our focus is on understanding how the error originating from the charm threshold region in \cref{momentsintegral}  propagates to  the extraction of the quark mass using the moments method. From \cref{momentspert,momentsintegral} and error propagation we obtain,  
\begin{equation}
     \frac{\delta\hat{m}_c}{\hat{m}_c}=\frac{1}{2n}\frac{\left(2\hat{m}_c\right)^{2n}}{\hat{C}_n}\delta\left[\int^{\left(5.0\,\mathrm{GeV}\right)^2}_{\left(3.7\,\mathrm{GeV}\right)^2}\frac{ds}{s^{n+1}}R^{\mathrm{cont}}_c\right],\label{errorprop}
\end{equation}
where $\delta$ denotes error. As an example let us use the second moment $n=2$ ($\hat{C}_2=0.63$), used by \cite{Erler:2016atg}. From \cref{charmlowint,errorprop}, we obtain $\delta \hat{m}_c<0.010\,\mathrm{GeV}$. The smaller contribution to the error from the threshold region, compared to the one shown in  \cref{charmmasscompared} is a consequence of the stronger suppression for large $s$ in the integral. Essentially, by obtaining the charm quark mass from the moments and using it as an input in the computation of $\Delta\alpha$ rather than explicitly integrating over the kernel of $\Delta\alpha$, the error is significantly reduced. This is because the zeroth moment is mainly sensitive to logarithmic dependence on the mass, leading to a bad precision. We also point out that our findings are consistent  with Ref.~\cite{Erler:2016atg} where  higher values of the mass correspond to a lower moment,  see \cref{fig:moments}.

Another point in favor of the RGE method is that the input parameters $\hat{m}_c$ and $\hat{\alpha}_s$ can also be obtained from lattice QCD. For example, the last FLAG review \cite{FlavourLatticeAveragingGroupFLAG:2021npn}  reports $\delta\hat{m}_c=0.005\,\mathrm{GeV}$ and $\delta\hat{m}_c=0.013\,\mathrm{GeV}$ using $n_f=2+1$ and $n_f=2+1+1$ active quarks in the sea respectively. The same reference quotes $\delta\hat{\alpha}_s(M_Z)=0.0008$. Taking the smallest parametric errors, and using  $R$ data below $1.8\,\mathrm{GeV}$ as input in the RGE method would leave us with $\delta\Delta\alpha^{(5)}(M^2_Z)\sim 0.6\times10^{-4}$. If on the other hand, one uses this small parametric error and takes lattice as an input, it is possible to reach an error of 
$\delta\Delta\alpha^{(5)}(M^2_Z)\sim 0.4\times10^{-4}$.

Finally we see a tension between lattice and $R$ data, as the reader can see in \cref{fig:comparison_low3,fig:Dalpha5all}. This tension is also seen in the original lattice calculation of Ref. \cite{Ce:2022eix}  where a different low energy comparison of $\Delta\alpha$ is done.

\section{Conclusions}

In this work, we computed a theoretically driven result for $\Delta\alpha^{(5)}(M^2_Z)$ using lattice QCD and perturbative QCD. Our main result corresponds to the RGE method and is given in \cref{eq:RGElattice}. Additionally, we provide a simple perturbative formula for relating lattice calculations to the integral over $R$ at low energies in \cref{eq:conversionto lattice}. Our work allows us to see what improvements need to be made to increase the precision of $\Delta\alpha^{(5)}(M^2_Z)$.

For example, when results from the Mainz collaboration are used as low-energy input, we find that the dominant error comes from the low-energy lattice data, while the second source of error is parametric. On the other hand, when BMW results are used, the dominant error is parametric, with the lattice error being the second most relevant. In our results, we used a conservative error: $\delta\hat{\alpha}_s = 0.0016$. If instead, the error is taken from the FLAG review or the PDF average, the total error on $\delta\Delta\alpha^{(5)}(M^2_Z)$ can be as small as $0.4\times10^{-4}$, making this the most precise determination of $\Delta\alpha^{(5)}(M^2_Z)$ to date, already at the level required for future $e^{+}e^{-}$ high-energy colliders \cite{Jegerlehner:2019lxt}.

There are important consequences of our results in electroweak precision physics. For example, it is easier to take into account the parametric errors in a global fit. Furthermore, it allows for a straightforward inclusion of the correlation between the anomalous magnetic moment of the muon and $\Delta\alpha^{(5)}(M^2_Z)$. We recall that the large value of $\Delta\alpha^{(5)}(M^2_Z)$ when using lattice data as input will reduce the value of the SM prediction for the W boson mass, increasing the tension with experimental measurements. We leave such a global fit analysis for future work.

Here, we also analyzed different frameworks and methods for computing the contributions to $\Delta\alpha^{(5)}(M^2_Z)$ using cross-section data as input. There is good agreement between the three methods we examined - explicit integration of $R$ in the timelike region, the Euclidean split technique, and RGE - although the RGE and Euclidean split technique have smaller errors. The smaller error can be attributed to the way the charm quark contributions are handled. Rather than directly integrating the $\Delta\alpha^{(5)}(M^2_Z)$ kernel over the data, we found that a smaller error is achieved when the data is used to calculate the charm quark mass from the heavy quark moments first. Furthermore, we found tension between lattice and cross-section data, which, within errors, is consistent with the findings\footnote{See Fig. 12 of Ref. \cite{Ce:2022eix}.} of Ref. \cite{Ce:2022eix}. Further work is required to understand the origin of the differences. 

In summary, when it comes to computing $\Delta\alpha^{(5)}(M^2_Z)$ from low-energy data, we believe the RGE method is more convenient for several reasons. Firstly, it allows for systematic resummation of logarithms. Additionally, the higher order $\hat{\alpha}_s$ corrections are simpler and don't require the full mass dependence of the vacuum polarization function.

While this manuscript was being prepared for submission, a study about the interplay between lattice and cross-section data was published \cite{Davier:2023cyp}, where it is suggested that an increase of $\sim 5\%$ in the cross-section data around the $\rho$ peak region may explain discrepancies between them.

\section*{Acknowledgements}
We would like to sincerely thank Hubert Spiesberger and Mikhail Gorshteyn for their thoughtful review of the manuscript and their meaningful contributions. Additionally, we express our gratitude to Hartmut Wittig for providing valuable insights and engaging in enlightening conversations with us.
\pagebreak

\appendix

\section{Double bubble contributions}
In the following  $m_q\equiv\hat{m}_q(\hat{m}_q)$. The double bubble contribution is
\begin{equation}
\int^{\mu^2}_{0}\frac{R^{db}_{l}(s)}{s}ds=N_c\sum_{l}Q^2_{l}\left[\hat{a}^2_{s}A_2+\hat{a}^3_{s}A_3\right],
\end{equation}
$A_2$ is the same for  for both, charm and bottom quarks ($x=\frac{\mu^2}{m_q^2}$) 
\begin{equation}
A_2=x\left(\frac{2}{25}-\frac{2}{135} \ln x\right)+x^2\left(-\frac{1513 }{2116800}+\frac{\ln x}{5040}\right)+x^3\left(\frac{1853}{80372250}-\frac{ \ln x}{127575}\right),
\end{equation}
while for $A_3$ we only consider the charm quark effects,
\begin{equation}
A_3=\frac{\mu^2}{m_c^2}\left(
         -\frac{349 \ln ^2\left( \frac{\mu^2}{m_c^2}\right)}{4860}+\frac{349051 \ln \left(\frac{\mu^2}{m_c^2}\right)}{874800}-\frac{1847 \zeta_3}{19440}+\frac{5 \pi ^2}{2916}-\frac{2585137}{26244000}\right).
\end{equation}

\section{RGE}
The coefficients mentioned in \cref{eq:fine structure} are: 
\begin{eqnarray}
\nonumber
K_q &=& N_c\left\{ 1 + \frac{3}{4} Q_q^2 \frac{\hat\alpha}{\pi} + \frac{\hat\alpha_s}{\pi} + 
\frac{\hat\alpha_s^2}{\pi^2} \left[ \frac{125}{48} - \frac{11}{72} n_q \right] \right. \\[12pt]
\nonumber
&+& \frac{\hat\alpha_s^3}{\pi^3} \left[ \frac{10487}{1728} + \frac{55}{18} \zeta_3 - 
n_q \left( \frac{707}{864} + \frac{55}{54} \zeta_3 \right)  - \frac{77}{3888} n_q^2 \right] \\[12pt] 
\nonumber 
&+& \frac{\hat\alpha_s^4}{4\pi^4} \left[ \frac{2665349}{41472} + \frac{182335}{864}\zeta_3 - 
\frac{605}{16}\zeta_4 - \frac{31375}{288}\zeta_5 \right. \\[12pt]
\nonumber
&-& n_q \left( \frac{11785}{648} + \frac{58625}{864}\zeta_3 - \frac{715}{48}\zeta_4 - 
\frac{13325}{432}\zeta_5 \right) \\[12pt]
&-& n_q^2 \left( \frac{4729}{31104} - \frac{3163}{1296}\zeta_3 + \frac{55}{72}\zeta_4 \right)  
+ n_q^3 \left.\left. \left( \frac{107}{15552}+\frac{1}{108}\zeta_{3}\right) \right] \right\},
\label{eq:ki}
\end{eqnarray}

and
\begin{eqnarray}
\nonumber
\sigma = \frac{\hat\alpha_s^3}{\pi^3} \left[ \frac{55}{216} - \frac{5}{9}\zeta_3 \right]
+ \frac{\hat\alpha_s^4}{\pi^4} \left[ \frac{11065}{3456} - \frac{34775}{3456}\zeta_3 + 
\frac{55}{32}\zeta_4 + \frac{3875}{864}\zeta_5 \right. \\[12pt]
- \left. n_q \left( \frac{275}{1728} - \frac{205}{576}\zeta_3 + \frac{5}{48}\zeta_4 + 
\frac{25}{144}\zeta_5 \right)\right].
\label{eq:sigma}
\end{eqnarray}
\subsection{Solution}
As seen in the text, the RGEs for $\hat{a}$ and $\hat{a}_s$ can be written as ($L\equiv\ln\mu^2$) \begin{equation}
\frac{d\hat{a}}{dL}=\hat{a}^2\beta,\quad\quad\quad\quad\quad\frac{d\hat{a}_s}{dL}=\hat{a}_s^2\beta_{QCD}.\label{RGEgen}
\end{equation}
where  \begin{equation}
\beta=\beta^{(0)}+\hat{a}\beta^{(1)}+\sum_{n}\delta_n \hat{a}^n_s,\label{expansionRGE}
\end{equation}
 \begin{equation}
\beta_{QCD}=\sum_{k}\beta^{(k)}_{QCD}\hat{a}^k_s,
\end{equation}
where we have neglected the mixed term of the form  $\hat{a}\hat{a}_s$. Dividing \cref{RGEgen} by $\hat{a}^2$ and integrating from the scale $\mu_1$ to the scale $\mu_2$, we get 
\begin{equation}
\frac{1}{\hat{a}_1}-\frac{1}{\hat{a}_2}=\beta^{(0)}\ln\frac{\mu^2_2}{\mu^2_1}+\int \hat{a}\beta^{(1)} dL+\sum_{n}\delta_n \int \hat{a}^n_s dL,\label{sol1}
\end{equation}
where we have used \cref{expansionRGE}. The second term on the right hand side of \cref{sol1} can be written as \begin{equation}
\int \hat{a}\beta^{(1)}dL=\int \hat{a}\beta^{(1)} \frac{dL}{d\hat{a}}d\hat{a}\approx\frac{\beta^{(1)} }{\beta^{(0)} }\int \frac{1}{\hat{a}}d\hat{a}=\frac{\beta^{(1)} }{\beta^{(0)} }\ln\frac{\hat{a}_2}{{a}_1}\approx \frac{\beta^{(1)} }{\beta^{(0)} }\ln\left[\frac{1}{1-\hat{a}_1\beta^{(0)}\ln\frac{\mu^2_2}{\mu^2_1}}\right],
\end{equation}
similarly, \begin{equation}
\int \hat{a}^n_s dL=\int \hat{a}^n_s \frac{dL}{d\hat{a}_s}d\hat{a}_s=\int \frac{\hat{a}^{(n-2)}_s}{\sum_k\beta^{(k)}_{QCD}\hat{a}^k_s}d\hat{a}_s.
\end{equation}
The integrand can be Taylor expanded, which allows to perform the integration analytically.  Hence we may rewrite \cref{RGEgen} as \begin{equation}
\frac{1}{\hat{a}_2}=\frac{1}{\hat{a}_1}-\beta^{(0)}\ln\frac{\mu^2_2}{\mu^2_1}-\frac{\beta^{(1)}}{\beta^{(0)}_{QED} }\ln\left[\frac{1}{1-\hat{a}_1\beta^{(0)}_{QED}\ln\frac{\mu^2_2}{\mu^2_1}}\right]+c_0\ln\frac{\hat{a}_s(\mu_2)}{\hat{a}_s(\mu_1)}+\sum_{n=1}c_n\left[\hat{a}^n_s(\mu_2)-\hat{a}^n_s(\mu_1)\right],
\end{equation}
where  \begin{equation}
c_0=\tilde{\delta}_1,\quad\quad\quad c_1=\tilde{\delta}_2-\tilde{b}_1 \tilde{\delta}_1, \quad\quad\quad
c_2=\frac{1}{2} \left( \tilde{b}_1^2 \tilde{\delta}_1-\tilde{b}_1 \tilde{\delta}_2-\tilde{b}_2
   \tilde{\delta}_1+\tilde{\delta}_3\right),
\end{equation}
\begin{equation}
c_3=\frac{1}{3} \left(-\tilde{b}_1^3 \tilde{\delta}_1+\tilde{b}_1^2 \tilde{\delta}_2+2 \tilde{b}_1 \tilde{b}_2
   \tilde{\delta}_1-\tilde{b}_1 \tilde{\delta}_3-\tilde{b}_2 \tilde{\delta}_2-\tilde{b}_3 \tilde{\delta}_1+\tilde{\delta}_4\right),
\end{equation}
\begin{equation}
4 c_4=\tilde{b}_1^4 \tilde{\delta}_1-\tilde{b}_1^3 \tilde{\delta}_2-3 \tilde{b}_1^2 \tilde{b}_2
   \tilde{\delta}_1+\tilde{b}_1^2 \tilde{\delta}_3+2 \tilde{b}_1 \tilde{b}_2 \tilde{\delta}_2+2 \tilde{b}_1 \tilde{b}_3
   \tilde{\delta}_1-\tilde{b}_1 \tilde{\delta}_4+\tilde{b}_2^2 \tilde{\delta}_1-\tilde{b}_2 \tilde{\delta}_3-\tilde{b}_3
   \tilde{\delta}_2-\tilde{b}_4 \tilde{\delta}_1+\tilde{\delta}_5 .
\end{equation}
and $\tilde{\delta}_n=\delta_n/\beta^{(0)}_{QCD}$, $\tilde{b}_n=\beta^{(n)}_{QCD}/\beta^{(0)}_{QCD}$. This  allows for a numerical implementation, computationally efficient for global fits.
\section{Matching conditions}
Here we state the results of \cite{Sturm:2014nva} with the group factors simplified at the scale $\mu=\hat{m}_q(\hat{m}_q)$, the logs may be easily reincorporated through the RGE of $\hat{\alpha}$.   \begin{align}
\Delta\hat{\alpha}^{(n_q-1)}(\hat{m}^2_q)&= \Delta\hat{\alpha}^{(n_q)}(\hat{m}^2_q)-\frac{15}{16}N_c Q^4_qa^2\left(1+\Delta\hat{\alpha}^{(n_q)}(\hat{m}_q)\right)-a Q^2_q\left\{\hat{a}^{(n_q)}_s\frac{13}{12}\right.\, \nonumber\\[12pt]
&+ \hat{a}^{(n_q)\,2}_s \left[ \frac{361}{1296}n_q+\frac{655}{144}\zeta_3-\frac{3847}{864} \right] \, \nonumber\\[12pt]
&+ \hat{a}^{(n_q)\,3}_s \left[ -\frac{85637 a_4}{1620}-\frac{656 a_5}{27}-\frac{928399
   \zeta_2^2}{129600}-\frac{1289}{135} \zeta_2^2 l_2-\frac{164}{81} \zeta_2
   l^3_2 \right.\, \nonumber\\[12pt]
&+  \left. \frac{85637 \zeta_2 l^2_2}{6480}-\frac{49 \zeta_5}{32}+\frac{42223463 \zeta_3}{604800}-\frac{321165301}{21772800}+\frac{82 l^5_2}{405}-\frac{85637 l^4_2}{38880} \right. \, \nonumber\\[12pt]
&+ n_q \left( -\frac{17 a_4}{27}+\frac{4487 \zeta^2_2}{2160}+\frac{17}{108} \zeta_2 l^2_2-\frac{21379 \zeta_3}{5184}-\frac{86101}{62208}-\frac{17 l^4_2}{648}\right) \, \nonumber\\[12pt]
&+ n^2_q\left.\left.\left(\frac{17897}{93312}-\frac{31}{216} \zeta_3\right)\right]\right\}-a\sum_{(l \neq f)}Q^2_l\left\{\hat{a}^{(n_q)\,2}_s\frac{295}{1296}\right.
\, \nonumber\\[12pt]
&+ \hat{a}^{(n_q)\,3}_s\left[\frac{67}{360} \zeta^2_2+\frac{1}{9}\zeta_2 l^2_2+\frac{163}{162}\zeta_3-\frac{86369}{186624}-\frac{l_2^4}{54}-\frac{4 a_4}{9}\right.
   \, \nonumber\\[12pt]
&+ \left.\left.\left(\frac{6625
   }{46656}-\frac{11\zeta_3}{108}\right)n_q\right]\right\}-aQ^2_q\hat{a}^{(n_q)\,3}_s\left\{
\frac{2411}{6048}-\frac{365 a_4}{36}
\right.  \, \nonumber\\[12pt]
&+\left.\frac{2189\zeta^2_2}{576}+\frac{365}{144}\zeta_2 l^2_2-\frac{25 \zeta_5}{72}-\frac{6779 \zeta_3}{1344}-\frac{365
   l^4_2}{864} \right\}
  \, \nonumber\\[12pt]
&- a\sum_{l\neq f}Q_qQ_l\hat{a}^{(n_q)\,3}_s\left\{
-\frac{\zeta^2_2}{6}-\frac{25 \zeta_5}{36}+\frac{655 \zeta_3}{432}+\frac{515}{1296}
\right\},
\label{matchingconditions}
\end{align}
where $a_n=\mathrm{Li}_n(1/2)$, $l_2=\ln(2)$ and $N_c$ is the number of colors. The term proportional to $Q^2_l$ corresponds to the contribution from double bubble diagrams, where the heavy quark bubble is inside light quark one. The terms proportional to $Q^2_q$ and $Q_l Q_q$  come from OZI violating diagrams, \textit{i.e} from disconnected double bubble diagrams, the so-called singlet piece. The same formula applies for leptons, making $\hat{a}_s=0$ and $N_c=1$.

\bibliography{apssamp}
\bibliographystyle{JHEP}

\end{document}